\documentclass[a4paper,12pt]{extarticle}
\usepackage[margin=3cm]{geometry}
\usepackage{newtxtext}
\usepackage[slantedGreek]{newtxmath}
\usepackage{natbib}
\usepackage{url}
\DeclareUrlCommand\doi{\urlstyle{rm}}
\usepackage{hawkingpapers}
\title{Stephen William Hawking CH CBE\\
8 January 1942 -- 14 March 2018\\
Elected FRS 1974}

\author{Bernard~J.~Carr$^{1}$, George~F.~R.~ Ellis$^{2}$, Gary~W.~Gibbons$^{3}$, James~B.~Hartle$^{4}$,\\
Thomas~Hertog$^{5}$, Roger~Penrose$^{6}$, Malcolm~J.~Perry$^{3}$, and Kip~S.~Thorne$^{7}$.\footnote{The editor-in-chief is most grateful to the authors for providing their contributions about the areas in which they worked with Stephen.  The Memoir was put together and edited by Malcolm Longair and Martin Rees in consultation with the authors.}\\
\\
$^1$ \emph{\small Queen Mary College, University of London, Mile End Road, London, E1 4NS, UK}\\[6pt]
$^2$ \emph{\small University of Cape Town, Rondebosch, Cape Town, 7700, South Africa}\\[6pt]
$^3$ \emph{\small DAMTP, University of Cambridge, Wilberforce Road, Cambridge CB3 0WA, UK}\\[6pt]
$^4$ \emph{\small Department of Physics, University of California, Santa Barbara,  93106, USA}\\[6pt]
$^5$ \emph{\small Institute for Theoretical Physics, KU Leuven, 3001 Leuven, Belgium}\\[6pt]
$^6$ \emph{\small Mathematical Institute, University of Oxford, Andrew Wiles Building,}\\
\emph{\small Radcliffe Observatory Quarter, Woodstock Road, Oxford OX2 6GG, UK}\\[6pt]
$^7$ \emph{\small Theoretical Astrophysics, California Institute of Technology, Pasadena, California 91125, USA}
}

\date{20 December 2018}
\begin{document}

\maketitle

\begin{abstract}
Stephen Hawking's contributions to the understanding of gravity, black holes and cosmology were truly immense.  They began with the singularity theorems in the 1960s followed by his discovery that black holes have an entropy and consequently a finite temperature.  Black holes were predicted to emit thermal radiation, what is now called Hawking radiation.  He pioneered the study of primordial black holes and their potential role in cosmology.  His organisation of and contributions to the Nuffield Workshop in 1982 consolidated the picture that the large-scale structure of the universe originated as quantum fluctuations during the inflationary era.  Work on the interplay between quantum mechanics and general relativity resulted in his formulation of the concept of the wavefunction of the universe.  The tension between quantum mechanics and general relativity led to his struggles with the information paradox concerning deep connections between these fundamental areas of physics.

These achievement were all accomplished following the diagnosis during the early years of Stephen's studies as a post-graduate student in Cambridge that he had incurable motor neuron disease -- he was given two years to live. Against all the odds, he lived a further 55 years.   The distinction of his work led to many honours and he became a major public figure, promoting with passion the needs of disabled people.  His popular best-selling book \emph{A Brief History of Time} made cosmology and his own work known to the general public worldwide.  He became an icon for science and an inspiration to all.
\end{abstract}

\section*{Foreword}

\begin{quotation}
`The image of Stephen Hawking in his motorised wheelchair, with head contorted slightly to one side and hands crossed over to work the controls, caught the public imagination, as a true symbol of the triumph of mind over matter. As with the Delphic oracle of ancient Greece, physical impairment seemed compensated by almost supernatural gifts, which allowed his mind to roam the universe freely, upon occasion enigmatically revealing some of its secrets hidden from ordinary mortal view. Of course, such a romanticised image can represent but a partial truth. Those who knew Hawking would clearly appreciate the dominating presence of a real human being, with an enormous zest for life, great humour, and tremendous determination, yet with normal human weaknesses, as well as his more obvious strengths.  He was extremely highly regarded, in view of his many greatly impressive, sometimes revolutionary, contributions to the understanding of the physics and the geometry of the universe.'
\end{quotation}
\qquad\qquad\qquad Roger Penrose \citeyearpar{penrose2018}.

\begin{quotation}
`We remember Newton for answers. We remember Hawking for questions. And Hawking's questions themselves keep on giving, generating breakthroughs decades later. When ultimately we master the quantum gravity laws, and fully comprehend the birth of our universe, it will be by standing on the shoulders of Hawking.'
\end{quotation}
\qquad\qquad\qquad Kip Thorne \citeyearpar{thorne2018}.

\begin{quotation}
`Few, if any, of Einstein's successors have done more to deepen our insights into gravity, space and time.'
\end{quotation}
\qquad\qquad\qquad Martin Rees \citeyearpar{rees2018}.

\pagebreak

\part{LIFE}

\section{Early life}

Stephen was born in Oxford on 8 January 1942, the 300th anniversary of the death of Galileo Galilei (Figure 1).\footnote{Many more details about Stephen's life can be found in the book \emph{Stephen Hawking: His Life and Work} \citeyearpar{Ferguson2017} by Kitty Ferguson.} His father, Frank Hawking, came from a family of tenant farmers in Yorkshire who suffered hard times during the agricultural depression at the beginning of the twentieth century.  Although financially stretched, the family was able to send Frank to Oxford where he studied medicine. His research expertise was in tropical medicine which involved regular field trips to East Africa.   At the beginning of the Second World War, despite volunteering for military service, the authorities judged that it would be best if Frank continued his medical research during the war years.  Stephen's mother, Isobel Walker, was born in Dunfermline in Scotland, but the family moved to Devon when she was twelve.  Isobel gained entrance to Oxford University where she studied economics, politics and philosophy.  She then worked for the Inland Revenue but this proved not to be to her taste and she subsequently became a school teacher.  She was a free-thinking radical and a strong influence on her son.

The family lived in London but, because of the threat of bombing, his mother moved to Oxford where Stephen was born.  He had two younger sisters, Mary born in 1943 and Philippa in 1947.  The family moved from Highgate in London to St.\,Albans in 1950 when Frank took up a post at the new National Institute for Medical Research at Mill Hill.    Stephen had all the enthusiasms of an enquiring young boy -- model trains, both clockwork and electric, boats and model aircraft as well as the invention of very complicated games with his close school friends Roger Ferneyhough and John McClenahan.

In Highgate, Stephen was educated at the progressive Byron House School.  When the family moved to St.\,Albans he attended the St.\,Albans School for Girls, which also took boys up to the age of ten (Figure 2).  As a consequence, his eleven-plus examinations were taken successfully a year early and he entered the boy's school, St.\,Albans School, in the  top stream.  Generally, he was in the middle of the class but obtained a good education with talented school-fellows.  His classmates gave him the prophetic nickname `Einstein'.

Frank was dedicated to research in tropical diseases and very hard-working.  He strongly encouraged Stephen's interest in science, taking him to his laboratory at Mill Hill to peer through microscopes and visit the insect house where mosquitos infected with tropical diseases were kept.  Frank also coached Stephen in mathematics until he could not keep up with Stephen in mathematical knowledge and skill.

In the last two years at high school, Stephen concentrated upon mathematics and physics and was fortunate in having a brilliant mathematics teacher Dikran Tahta.  With Tahta's help, Stephen built an early primitive computer.  Later, Stephen stated that `Thanks to Mr Tahta, I became a professor of mathematics at Cambridge, a position once held by Isaac Newton.' \citep{Ferguson2017}

Frank was keen that Stephen should try to gain entrance to University College, Frank's old College at Oxford University.  Although the headmaster at St.\,Albans thought Stephen too young to take the entrance examination, he succeeded at the age of 17.

\section{Oxford}

Most of Stephen's colleagues at Oxford were older than him, some having done military service.  In his Memoirs Stephen records being somewhat lonely in his first two years and so joined the College boat club as a cox in his third year (Figure 3) (\ref{Hawking2013}).  Although he was not a distinguished cox, the experience expanded his circle of friends.  At that time, the only examination which mattered in Oxford was the final third year examination.  As he wrote, the physics course was designed in such a way that it was easy to avoid work. Furthermore:
\begin{quote}
`To work hard to get a better class of degree was regarded as the mark of a ``grey man'', the worst epithet in the Oxford vocabulary.'
\end{quote}
Stephen did not work hard during his Oxford years, although his talent was recognised.  For the final examinations he concentrated upon problems in theoretical physics which did not require much factual knowledge.  By cramming before the examination, he attained a borderline first--second class degree.  To judge which degree should be awarded, an interview with the examiners was held.  Stephen told them that, if he got a first, he would go to Cambridge to do research -- if awarded a second, he would stay in Oxford.  He was awarded a first.

\section{Cambridge}

The 1960s were an exciting period in astronomy, astrophysics and cosmology.  It is no exaggeration to say that the discoveries and innovations of that decade changed these disciplines out of all recognition -- evidence began to emerge for the existence of black holes and the consolidation of the Big Bang as the preferred model for the large scale structure, dynamics and evolution of the universe.  Stephen seized the opportunities to create a remarkable body of original work which is described in Part II of this Memoir.

After an adventurous trip through Persia, which included being close to the epicentre of a 7.1 magnitude earthquake that killed more than 12,000 people, Stephen arrived at Trinity Hall, Cambridge in the autumn of 1962, planning to study under the distinguished astrophysicist and cosmologist Fred Hoyle (FRS 1957). Disappointingly for Stephen, Hoyle was unable to take him on.  The other person available in the cosmological field was Dennis Sciama (FRS 1983), who was unknown to Stephen at the time. In fact, this proved to be a piece of great good fortune, since Sciama was an outstandingly stimulating figure in British cosmology, supervising many students who were to make impressive names for themselves in later years.  During the years 1964--1973, when Sciama was a member of the Department of Applied Mathematics and Theoretical Physics (DAMTP), his graduate students included, in date order, George Ellis (FRS 2007), Brandon Carter (FRS 1981), Ray McLenaghan, Martin Rees (FRS 1979, PRS 2005), John Stewart, Malcolm MacCullum, Bill Saslaw, Gary Gibbons (FRS 1999) and Bernard Jones.  Ellis and Penrose \citeyearpar{ellispenrose2010} describe Sciama's remarkable record.  Sciama seemed to know everything that was going on in physics at the time, especially in cosmology, and conveyed an infectious excitement to all who encountered him. He was also very effective in bringing together people who would benefit significantly from communicating with one another.

Already in his last year at Oxford, Stephen noticed that he was becoming increasing clumsy and sought medical advice.  His condition continued to decline following his move to Cambridge.  During the Christmas break in 1962, he fell over while skating at St.\,Albans and could not get up. Shortly after his 21st birthday a month later, Stephen was diagnosed as suffering from an unspecified incurable disease, which was later identified as the fatal degenerative motor neurone disease, amyotrophic lateral sclerosis (ALS).  The doctors' prognosis was that he probably had only two years to live.

While in hospital soon after his illness was first diagnosed, Stephen's depression was somewhat lifted when he compared his lot with that of a boy he vaguely knew in the next bed who was dying of leukemia.  Stephen resolved to do something really creative with his remaining years and aspired to tackle some of the most fundamental questions concerning the physical nature of the universe.   For the first time in his life he worked seriously hard and found he really enjoyed it -- he began to exploit his remarkable gifts in a series of revolutionary papers in gravitational physics.

Even more important for Stephen was his engagement in 1964 to Jane Wilde. She had grown up in St.\,Albans and studied languages at the University of London's Westfield College.  They met through mutual college friends at a party in 1962 and were married in 1965 in the chapel of Trinity Hall, Stephen's Cambridge College.  They had three children: Robert, born in 1967, Lucy in 1970 and Timothy in 1979 (Figure 4). She received her PhD in medieval Spanish poetry in April 1981.  Jane was exceptionally supportive of Stephen as his condition deteriorated.   Perhaps one of Jane's most important contributions was to allow Stephen, at his own insistence, to do things for himself to an unusual extent -- he was an extraordinarily determined person.  His health worsened further, until by the late 1970s, he had almost no movement left, and his speech could barely be made out at all, except by a very few of his family and colleagues.  Nonetheless, defying established medical opinion, he lived another 40 years.

In 1964, Stephen needed a job to support a family.  The originality of his work soon resulted in a succession of positions.  The first step on the ladder was a Research Fellowship of Gonville and Caius College, Cambridge which was to remain his college for the rest of his life.  This was followed by an appointment as a staff member of the Institute of Theoretical Astronomy during the most exciting period of its existence from 1968 to 1972.  In 1969 he was elected to full fellowship of Gonville and Caius College for Distinction in Science.   After the creation the Institute of Astronomy in 1972, Stephen remained there as a research assistant for two years before gaining a more permanent status at DAMTP.  By this time, the originality and importance of Stephen's work was recognised world-wide and he was elected a Fellow of the Royal Society in 1974 at the exceptionally early age of 32.  He would receive the Society's highest honour, the Copley medal, in 2006.

Stephen and the family visited CalTech in Pasadena for the academic year 1974 to 1975 as a Sherman Fairchild Distinguished Scholar, CalTech's highest award. He found Caltech and California exhilarating.  At that time, the facilities were much better than in Cambridge and there were ramps everywhere for his wheelchair, installed for the community at his behest.  Among the important events of the visit, Stephen gave a major seminar about Hawking radiation at Caltech in the presence of Richard Feynman.  During the year at Caltech, Stephen was awarded the Pius XI medal and flew to Rome to receive it. He insisted on visiting the Vatican archives to read the recantation of Galileo, with whom he always felt a strong affinity.

This was the beginning of Stephen's long-term relationship with Caltech. From 1991 through 2013, Stephen visited Caltech for several weeks nearly every year as a Fairchild Scholar.  From this base, his long-term research collaboration with Jim Hartle at the University of California, Santa Barbara, flourished and he developed close ties to Hollywood, which resulted in appearances on \emph{The Big Bang Theory}, \emph{Star Trek} and \emph{The Simpsons}.  These distinctive appearances on screen helped cement his role as a public icon for science.

During the early 1970s, he gradually lost the ability to write.  As Kip Thorne has described,
\begin{quotation}
`Stephen lost the use of his hands for writing equations in the early to mid 1970s, with the final, complete loss occurring during his 1974--75 year with me as a Fairchild Scholar at Caltech.  [Much of his research at that time] was in classical general relativity, and involved problems that could be cast in the language of geometry and topology.  As he lost the use of his hands, he developed an amazing ability to visualize and manipulate in his head geometric and topological concepts and relationships, and much of his breakthrough research relied on this.  It appears to me that his disability was a partial blessing in that it drove him to develop this ability to the point that it gave him insights that he might never have achieved otherwise.'
\end{quotation}

By this time, Stephen was so frail that most of his colleagues feared that he could scale no further heights.  But this was just the beginning.  As Martin Rees has written,
\begin{quotation}
`He worked in the same building as I did. I would often push his wheelchair into his office, and he would ask me to open an abstruse book on quantum theory, not a subject that had hitherto much interested him. He would sit hunched motionless for hours -- he couldn't even turn the pages without help. I wondered what was going through his mind, and if his powers were failing. But within a year, he came up with his best-ever idea -- encapsulated in an equation that he said he wanted inscribed on his memorial stone.' \citep{rees2018}
\end{quotation}
Latterly, students and colleagues would write a formula on a blackboard.  He would stare at it and say whether he agreed with it or not, and perhaps what should come next.

In 1979, Stephen was appointed to one of the most distinguished posts in the University as the 17th holder of the Lucasian Chair of Natural Philosophy, some 310 years after Isaac Newton (FRS 1672, PRS 1703) became its second holder. Stephen held this chair with distinction for 30 years until reaching the retirement age in 2009, after which he held a special research professorship, thanks to a generous endowment by the Avery-Tsui Foundation.  Dennis Avery and Sally Tsui Wong-Avery had earlier provided substantial support to the Stephen Hawking Centre for Theoretical Cosmology in DAMTP.

While in Switzerland in 1985, Stephen contracted pneumonia and a tracheotomy was necessary to save his life. Strangely, after this brush with death, the progression of his degenerative disease seemed to slow to a virtual halt. His tracheotomy prevented any form of speech so that acquiring a computerised speech synthesiser became a necessity. He was sustained, then and thereafter, by a team of helpers and personal assistants, as well as by the family.   In the aftermath of this encounter with pneumonia, the Hawkings' home was almost taken over by nurses and medical attendants, and he and Jane drifted apart.

They separated in 1990 and were divorced in 1995. In the same year, Stephen married Elaine Mason, who had been one of his nurses and whose former husband had designed Stephen's speech synthesiser.  Eventually, their relationship also came to an end -- they were divorced in 2007.  Stephen was supported, then and thereafter, by a team of helpers and personal assistants, as well as by his children and Jane.

\section{The Public Figure}

The feature film \emph{The Theory of Everything}, in which Stephen was astonishingly accurately portrayed in an Oscar-winning performance by Eddie Redmayne, recreated the human story behind his struggle. It surpasses most film biographies in representing the main characters sensitively and sympathetically, even though it understandably omitted, conflated and chronologically distorted key episodes in his personal and scientific life. It conveyed how the need for support, requiring a team of nurses, strained his marriage to Jane to breaking point, especially when augmented by the pressure of his growing celebrity.  Jane's book, \emph{Travelling to Infinity: My Life with Stephen}, on which the film is based, chronicles the 25 years during which, with amazing dedication, she underpinned their family life and his career \citep{hawkingJ2009}. Even before this film of 2014, Stephen's life and work had featured in an excellent TV docudrama made in 2004, in which he was played by Benedict Cumberbatch who also spoke his words in a 4-part documentary `The Grand Design' made for the Discovery TV Channel.

In his later life, Stephen became increasingly involved in the popularisation of science. This began with the astoundingly successful book \emph{A Brief History of Time} (1988), which was translated into some 35 languages and sold over 10 million copies in the following 20 years.   Undoubtedly, the book's brilliant title was a contributory factor to its success and the subject matter gripped the public imagination.   There is a directness and clarity of style, which Stephen developed as a matter of necessity when trying to cope with the limitations imposed by his physical disabilities. He could communicate only with great difficulty and expenditure of effort and so had to make do with short sentences that were directly to the point. In addition, one cannot deny that his physical condition powerfully caught the public's imagination.

Although the dissemination of science among a broader public was certainly one of Stephen's aims in writing the book, he also had the serious purpose of generating income. The financial needs were considerable, due to the demands of his entourage of family, nurses, healthcare helpers and increasingly expensive equipment. Some, but far from all, of this was covered by the National Health Service and grants.

To invite Stephen to a conference always involved serious challenges for the organisers. The travel and accommodation expenses were enormous, not least because of the sheer number of accompanying people.  Because of his fragile health, he had to fly first class and, in his last years, by private jet or air ambulance. But a popular lecture by Stephen would always be a sell-out, and special arrangements would be needed to find a large enough lecture hall.

In 1998 Stephen lectured at Clinton's White House and returned in 2009 when President Obama presented him with the US Medal of Freedom, a very rare honour for any foreigner (Figure 5).  This was just one of the many awards accumulated over his career, including the Companion of Honour from the UK (see the list of Honours and Distinctions). In the summer of 2012, he reached perhaps his largest-ever audience in his starring role at the opening ceremony of the London Paralympics.

Stephen clearly enjoyed his fame.  He welcomed opportunities to travel and to have unusual experiences. For instance, on a visit to Canada, he was undeterred by having to travel two miles down a mine-shaft to visit the underground Sudbury Neutrino Observatory (SNO). In 1997, the Chilean Air Force took a group of theoretical physicists including Stephen to its base on Isla Rey Jorge on the Antarctic Peninsula. Stephen remarked that, `Although my wheelchair has snow chains, they took me for a ride on a snow vehicle.'  He experienced weightlessness in NASA's reduced-gravity aircraft in 2007.

The presentational polish of his public lectures increased with the years.  In later years impressive computer-generated visuals were used while he controlled the verbal material, releasing it sentence by sentence in his computer-generated American-accented voice. High-quality pictures and computer-generated graphics also featured in his later popular books \emph{The Illustrated Brief History of Time} (1996) and \emph{The Universe in a Nutshell} (2001).  Stephen lent his support to his daughter Lucy when she wrote her series of six delightful childrens''s adventures in space, beginning with \emph{George's Secret Key to the Universe} (2007).  His last book published posthumously was entitled \emph{Brief Answers to the Big Questions} (2018).

Stephen enjoyed his work, the company of other scientists, trips to the theatre and the opera and his travels, which would have exhausted even a fully-fit academic (Figure 6). He took great pleasure in children, sometimes entertaining them by swivelling around in his motorised wheelchair. He could be generous and was often very witty.  But he also had a mischievous streak, examples including the wagers he made in the formal tradition of the senior combination room of Caius College's wager book. These included the denial of the presence of a black hole in Cygnus X-1, the prediction that the Higgs boson would not be found and, perhaps most significantly, that no information could return through a black hole event horizon -- he lost these three wagers.  On occasion he could display something of the arrogance that is not uncommon among physicists working at the cutting edge, as well as an autocratic streak. Yet he could also show a true humility.   Stephen continued, right until his last decade, to co-author technical papers, and speak at premier international conferences, doubly remarkable in a subject where even healthy researchers tend to peak at an early age.

Stephen supervised about 40 graduate students, some of whom later made significant names for themselves. Yet being a student of his was not easy. He was known to run his wheelchair over the foot of a student who caused him irritation. His pronouncements carried great authority, but his physical difficulties often caused them to be enigmatic in their brevity.  For the best students, he could provide instant inspiration.  As Bernard Carr recalls,
\begin{quotation}
`Students are probably always in awe of their supervisors and with Stephen the awe was even greater. Indeed, on matters of physics, I always regarded him as an oracle, just a few words from him yielding insights that would have taken weeks to work out on my own.'
\end{quotation}

Stephen had robust common sense, and was ready to express forceful political opinions. A downside of his iconic status was, however, that his comments attracted exaggerated attention even on topics where he had no special expertise -- for instance philosophy, or the dangers from aliens or from intelligent machines.

But there was absolutely no gainsaying his lifelong commitment to campaigns for the disabled and in his support of the UK National Health Service to which he owed so much.   He was firmly aligned with other political campaigns and causes.  When visiting Israel, he insisted on going also to the West Bank. Newspapers in 2006 showed remarkable pictures of Stephen, in his wheelchair, surrounded by fascinated and curious crowds in Ramallah.  Touchingly, just three months before his death, he gave a keynote address to a Sightsavers' event about his father's work on tropical medicine.

In an extraordinary demonstration of the clarity and the depth of his vision, Stephen transformed his insights in cosmology into a powerful message for humanity, which he promoted with wit and humour and with a missionary zeal. Of course cosmology cannot tell us how we should live. But Stephen felt it could provide a powerful overarching framework that emphasizes the unity behind it all and the truly gigantic potential lying ahead of us -- if only we can survive the boundaries of our time. He hoped to build a world in which humankind embraced with him that cosmic perspective to become guardians of our planet -- Spaceship Earth.
\begin{quote}
`Our only boundaries are the way we see ourselves',
\end{quote}
he said.  We must become global citizens, agents in the universe, and make sure the future will be a place we would like to visit.

\part{WORKS}

\section{Classical Gravitation Theory}

Stephen's best-known work on classical gravity is his set of singularity theorems.  This work concerned a hot topic at the time -- did the universe have a beginning?  More specifically, do initial singularities necessarily exist in cosmological models more general than the spatially homogeneous and spherically symmetric Robertson-Walker models, which have only one degree of freedom?  Jointly with Ellis, Stephen started with a proof in early 1965 of the existence of singularities in spatially homogeneous anisotropic world models (\ref{HawkingEllis1965}).  This was soon followed by the first paper in a series on singularity theorems which made use of new techniques, involving ideas from differential topology and space-time causal structure introduced by Roger Penrose (FRS 1972) in the context of gravitational collapse to a black hole. Penrose's theorem was based on his notion of a \emph{trapped surface}, which, without any appeal to symmetry, characterised a collapse that had reached a point of no return \citep{penrose1965}. Stephen's first contribution using these techniques was to notice that in an expanding cosmology there can be time-reversed trapped surfaces on a cosmological scale, so that Penrose's theorem could be applied immediately to demonstrate the existence of generic singularities in the past, in spatially non-compact expanding cosmologies (\ref{Hawking1965a}).

Stephen's result, along with Penrose's original theorem, required the existence of a global Cauchy hypersurface that is non-compact and, for cosmological considerations in particular, this requirement may be regarded as a serious limitation to the applicability of the theorem. Accordingly, in order to eliminate such restrictions, Stephen set his mind to developing new techniques that might strengthen the existing results. Concerted study of Morse theory and related topics in conventional Riemannian geometry was involved, and then developed for application in Lorentzian space-times. He was led to introduce notions such as Cauchy horizons and the concept of strong causality, which enabled him to prove new results on the generic occurrence of singularities in cosmological situations.  The results were published in a series of three remarkable papers in the \emph{Proceedings of the Royal Society} (\ref{Hawking1966c},\ref{Hawking1966e},\ref{Hawking1967}), also becoming Chapter 4 of his PhD dissertation.

Stephen's striking results showed that singularities are inevitable under very general circumstances where a plausible physical restriction on the energy-momentum tensor, referred to as the `strong energy condition', holds but there is no other requirement on the nature of the matter source. In the course of this highly innovative work, Stephen gained assistance from several colleagues, most notably from Brandon Carter, then a co-PhD student in DAMTP, who pointed out various technical errors in earlier drafts, all of which Stephen was able to circumvent.  In addition to colleagues mentioned above, Stephen acknowledged useful discussions with Sciama, Charles Misner, and Larry Shepley and the ideas continued to be developed in further papers (\ref{Hawking1966b},  \ref{Hawking1966d}, \ref{HawkingEllis1968}).  Finally, in a paper written jointly with Penrose in 1970, he used a new idea, developed independently by both authors, which provided a particularly general result, covering both cosmology and local collapse to a black hole, which encompassed almost all the results in these area that had gone before (\ref{HawkingPenrose1970}) (Figure 7).

In addition to this Stephen did other work in classical general relativity theory that is an important part of his legacy.  His very first paper concerned the Hoyle--Narlikar action at a distance theory of gravity of 1964, which he showed was not viable in expanding Robertson-Walker universes (\ref{Hawking1965}). This work was developed under the guidance and strong support of Sciama, his research supervisor. Stephen famously challenged Fred Hoyle about the Hoyle--Narlikar theory at a Royal Society meeting in London in June 1964, claiming that there were divergences in the theory in the context of an expanding universe when calculated as the sum of half the retarded and half the advanced fields, because the advanced field would be infinite. Hoyle asked ``How do you know?'', and Stephen answered ``Because I calculated it!''. This demonstrated already Stephen's independence of thought and refusal to be cowed, as a graduate student, by one of the foremost cosmologists of the day. This work formed Chapter 1 of Stephen's PhD dissertation.

A third theme concerned perturbations in the expanding universe. Stephen first used a transparent 1+3 covariant formalism developed by the Hamburg group of Otto Heckmann, Engelbert Sch\"ucking, J\"urgen Ehlers, Wolfgang Kundt, Rainer Sachs and Manfred Tr\"umper to examine the growth of density perturbations, rotational perturbations and the transmission and absorption of gravitational radiation in both standard cosmologies and the Steady State universe (\ref{Hawking1966a}).  The advantage of these covariant methods is that they avoid treating perturbations which are merely due to coordinate transformations as physical, providing a very clear view of gravitational wave propagation and absorption based on physically transparent variables. Stephen confirmed the well-known result that the predicted statistical density fluctuations were too small to create galaxies as observed. This work became Chapter 2 of his dissertation. His analysis, however, did not in fact define density perturbations in a gauge invariant way, but this problem was later remedied in a series of papers starting with that of Ellis and Marco Bruni \citeyearpar{ellisbruni1989} on the 1+3 covariant and gauge invariant approaches to perturbations in cosmology.

Stephen next used asymptotic expansions and the Newman--Penrose formalism to examine gravitational radiation in the expanding universe -- this became Chapter 3 of his dissertation.  Such radiation analyses in asymptotically flat universes were hot topics at the time. Stephen was assisted by co-PhD student Ray McLenaghan in the calculation of the Bianchi identities used in this study. Following Newman and Penrose, Stephen showed that `peeling' theorems established in the asymptotically flat case held also in more general scenarios in a slightly modified form, and derived the asymptotic symmetry group.  Unlike the asymptotically flat case, it is the same as the isometry group of the undisturbed space-time. A quantity was defined which was interpreted as the total mass of the source and the disturbance, and which monotonically decreases as gravitational radiation is emitted. These calculations were probably the most onerous in the whole body of Stephen's work (\ref{Hawking1968}).

The research up to this point was contained in Stephen's remarkable PhD dissertation, which was approved on 1 February 1966 (\ref{Hawking1966f}). It is a typed document with hand written equations, and erroneous parts crossed out by hand.   The abstract reads:
\begin{quotation}
`Some implications and consequences of the expansion of the universe are examined. In Chapter 1 it is shown that this expansion causes grave difficulties for the Hoyle--Narlikar theory of gravitation. Chapter 2 deals with perturbations of an expanding homogeneous and isotropic universe.  The conclusion is reached that galaxies cannot be formed as a result of perturbations that were initially small. The propagation and absorption of gravitational radiation is also investigated in this approximation. In Chapter 3 gravitational radiation in an expanding universe is examined by a method of asymptotic expansions. The `peeling off' behaviour and the asymptotic group are derived. Chapter 4 deals with the occurrence of singularities in cosmological models. It is shown that a singularity is inevitable provided that certain very general conditions are satisfied.'
\end{quotation}
The concise nature of this abstract clearly demonstrates his ability to focus on the essentials of the issue at hand.

Stephen next considered primordial nucleosynthesis in spatially homogeneous cosmologies in a paper with Roger Tayler (FRS 1995) (\ref{HawkingTayler1966}). They considered how anisotropies in cosmology could reduce helium production by primordial nucleosynthesis and so improve agreement with the (erroneous) helium-4 estimates of the time. The results depended on both density estimates and the length of time during which the anisotropy dominated, as pointed in a footnote by Thorne \citeyearpar{thorne1967}. Stephen then studied the properties of rotating Bianchi models (\ref{Hawking1969}) and, later with Barry Collins, the anisotropies in the Cosmic Microwave Background Radiation (CMB) in rotating Bianchi models (\ref{CollinsHawking1973a}, \ref{CollinsHawking1973b}), which they controversially related to the anthropic principle. The aim of these studies was to place strong limits on the rotation of the universe, and then to explain why that rotation was so small.

A further theme was classical black hole studies which started with the paper of 1971 (\ref{Hawking1971}) and which led to the area theorem, uniqueness theorems and the four laws of black hole thermodynamics, discussed in Section 7.

Stephen's work on singularities for the period up to December 1966 was consolidated in his Adams Prize essay (\ref{Hawking1966d}), recently republished with commentary \citep{ellis2014}. The prize that year was awarded to Roger Penrose, but Stephen was awarded an `auxiliary Adams Prize' for his essay \emph{Singularities and the structure of space-time}.  It has six sections, giving an outline of Riemannian geometry and general relativity, discussing the physical significance of curvature, the properties of causal relations, and some singularity theorems. Stephen states
\begin{quotation}
`Undoubtedly, the most important results are the theorems in Section 6 on the occurrence of singularities. These seem to imply either that the General Theory of Relativity breaks down or that there could be particles whose histories did not exist before (or after) a certain time. The author's own opinion is that the theory probably does break down, but only when quantum gravitational effects become important. This would not be expected to happen until the radius of curvature of spacetime became about $10^{-14}$ cm.'
\end{quotation}

The singularity work up to 1972, including partial classical black hole uniqueness theorems and the area theorem, were consolidated in the text \emph{The Large Scale Structure of Space Time} by Stephen and Ellis which from then on served as a standard reference in the field (\ref{HawkingEllis1973c}).  The writing began in 1970 when the Hawking--Penrose singularity theorem more or less cleared up the field, but there were many relevant results that were needed to comprehend what had been done and so Stephen and Ellis, both then post-doctoral fellows at DAMTP, decided to write this monograph to pull them all together.  While the writing was in progress, various results on causality and black hole theory appeared, including Stephen's crucial black hole area theorem, which had been inspired by earlier discussions with Penrose, and so substantial sections of the book were devoted to these topics.  Some important themes of classical black hole theory such as the Four Laws of Black Hole Thermodynamics and the definitive uniqueness theorems were not included, because they were developed later.

\section{Gravitational Radiation -- An Experimental Digression}

Meanwhile, in 1969 Joseph Weber reported the detection of gravitational radiation from the centre of our Galaxy.  This was a controversial result since, at the sensitivity level achievable with Weber's bar-type detector, the luminosity of the Galactic Centre would have been enormous.  There was scepticism that the result was real, but it needed independent checking.    Following a visit to DAMTP by Peter Aplin from Bristol University, Gibbons and Stephen started thinking about bar-type detectors for gravitational waves and their sensitivity.  They wrote a paper on the possibility of their detection (\ref{SWH25}) and then, encouraged by both Sciama and George Batchelor (FRS 1957), at the time the Head of DAMTP,  Stephen went so far as to make an application for funds to construct an improved detector which would be built in the basement of DAMTP. Other proposals were received from the Universities of Glasgow and Bristol.  Following a meeting among the various parties at the Science Research Council, the Cambridge proposal was withdrawn and the Glasgow proposal, led by Ron Drever and James Hough, was supported because of their technical expertise in experimental gravitation.    In the same year, Stephen published a paper on limits to the available energy in gravitational waves emitted by colliding black holes (\ref{SWH23}) (see Section 7).  He also published a review for non-experts of the theory of gravitational radiation (\ref{SWH27}).

There was a world-wide effort to check Weber's results which proved not to be reproducible, but the resulting activity stimulated the search for gravitational waves by a new generation of astronomical technologists.  This culminated in the success of the LIGO project about 45 years later with an increase in strain sensitivity from $\Delta L/L \approx 10^{-16}$ to $\Delta L/L \leq 10^{-23}$, over a factor $10^7$ in measurement precision.

With the development of the theory of inflation it became clear that gravitational waves should have been emitted during inflation as quantum fluctuations of a transient de Sitter phase and are potentially observable today through their polarisation signature in the intensity fluctuations of the Cosmic Microwave Background Radiation.  In 2000 Stephen returned to their study with Neil Turok and Thomas Hertog (\ref{SWH189}) using methods far removed from those of the paper of 1966 (\ref{Hawking1966a}) and closer in spirit to those developed in 1977 with Gibbons (\ref{SWH49}).

\section{Classical Black Hole Physics}

A distinct phase of Stephen's career concerned investigations of classical black hole physics.  The singularity theorems had clearly piqued his interest in  black holes as the Hawking--Penrose theorems include the result that a space-time that contains a marginally trapped surface must be singular. The result was interpreted as meaning that a black hole must contain a space-time singularity.  At the start of this phase, many of the fundamental physical properties of black holes were not well understood.

In 1965, Penrose had discovered the concept of the `apparent horizon' of a black hole, the point at which light and all matter are locally trapped, and he used this concept, differential topology and Einstein's equations to prove that in the core of every black hole there must be a physical singularity \citep{penrose1965}; this proof was later strengthened by Penrose and Hawking (\ref{HawkingPenrose1970}). Then Stephen, using Penrose's tools, discovered the crucial concept of the `event horizon', the surface about an evolving black hole at which light and matter are absolutely trapped and can never escape -- he showed that the event horizon always surrounds the apparent horizon (\ref{SWH23},\ref{SWH24}).

Stephen was concerned that black holes might never be detectable. Yakov Zel'dovich and Oktay Guseinov \citeyearpar{zeldovichguseynov66} had already suggested that natural places to look were in those binary systems in which there was an invisible companion heavier than the visible component. Stephen and Gibbons pointed out a refinement (\ref{SWH22}). In a binary system in which a black hole has formed, there is likely to be considerable mass loss from the partner that undergoes gravitational collapse. The collapse is therefore likely to make their orbits considerably eccentric, in contrast to binary systems that have been in equilibrium for a long time in which the eccentricity is generally expected to be small. Their idea was to look for black holes in highly eccentric systems. Their paper concluded with an examination of a small list of possible candidates and challenged the observational community to make more exhaustive studies of these systems.

The next paper, although clearly inspired by the study of gravitational radiation, is of much deeper significance (\ref{SWH23}). Suppose two black holes collide. Initially, the metric of each black hole is approximately given by the Kerr solution. Suppose that the areas of the event horizons of these black holes are $A_1$ and $A_2$. Stephen showed that the final state of these two black holes would be a more massive Kerr black hole with event horizon $A_3 \ge A_1+A_2$. This yields a bound on the amount of energy that the system can lose during the collision. As an example, suppose that the two initial black holes were of equal mass, $m$.  Then, the maximum amount of energy that could be lost from the system would be $(2 - {\sqrt 2})mc^2$. Consequently, the mass of the final black hole $M$ must obey the relation ${\sqrt 2}m \le M \le 2m$. When the discovery of gravitational radiation from the coalescence of two 30 $M_{\odot}$ black holes was announced in 2016, Stephen's first question was to ask if this bound was obeyed in the black hole collision -- he was overjoyed to hear that the observations were consistent with his expectation.

An elaboration of these ideas, together with two fundamental new results, was published in 1972 (\ref{SWH24}).  Previously, it had been assumed that a black hole in a stationary state was described by the Kerr metric and this new work went a long way to proving the conjecture correct. Firstly, Stephen showed that the horizon of a stationary black hole must be topologically spherical. Then he went on to prove the key result that, if a black hole is stationary but non-static, then it must be axisymmetric as is the case in Kerr metric.

There were competing theories of gravity. One that was popular at the time was the Brans--Dicke theory, which is a scalar-tensor theory that reduces to general relativity in the limit in which there is no scalar field. At the time there may have been some scepticism as to whether black holes could exist, simply because their properties seemed unpalatable to many.   Perhaps they did not exist in other viable theories of gravity.  Stephen showed that in the Brans--Dicke theory, black holes were the same as they are in general relativity (\ref{SWH26}).

Also in 1972, Hartle and Stephen answered a simple question -- is it possible to have more than one black hole in static equilibrium? (\ref{SWH29}) The answer is `Yes' if the black holes have electric charge. The idea is that the electrostatic repulsion between black holes can exactly cancel the gravitational attraction between them if their electric charge is equal to their mass in `natural' units. At the time, this just seemed like a trick but in fact it is one of the first examples of a Bogomol'nyi--Prasad--Sommerfield (BPS) state, which appears in supersymmetric theories. Although supersymmetry was unknown at the time, BPS states have a deep significance that has still to be completely understood.

In a further paper with Hartle, they revisited the exploration of the area of a black hole (\ref{SWH28}). The idea was to start with a Kerr black hole and ask how it responds to infalling matter or gravitational radiation. The principle result is that the area of the event horizon inevitably increases. There is an inescapable parallel with the second law of thermodynamics and at the time it was termed the second law of black hole mechanics, or sometimes just the area theorem.

Following the derivation of the area theorem, the early 1970s saw three crucial developments in classical black hole theory -- the cosmic censorship hypothesis, the `no hair' theorem and the laws of black hole mechanics, Stephen playing an important role in all of these.

The cosmic censorship hypothesis arose from the demonstration by Werner Israel (FRS 1986) in 1967 that unless the remnant of a non-rotating collapsing star was exactly spherical, the singularity it contained would be visible to outside observers \citep{PhysRev.164.1776}.  The breakdown of general relativity at such a `naked' singularity would then destroy the ability to predict the future of the rest of the universe. Penrose and John Wheeler therefore suggested that the remnant from the collapse of a non-rotating star would rapidly settle down to a spherical state, in which the singularity could not be seen from the outside the black hole \citep{1969NCimR...1..252P}. The idea was extended to the rotating case by the development of the no hair theorem. Together with Carter and David Robinson, Stephen showed that, once a black hole has settled down to equilibrium, it must be described by the solution of general relativity discovered by Roy Kerr \citeyearpar{1963PhRvL..11..237K}. Consequently, classical black holes are described by just two numbers, their mass and angular momentum. Later the no hair theorem was extended to include electric charge with the Kerr--Newman solution \citep{1965JMP.....6..918N}.  The no hair theorem now states that a black hole is described by just three parameters (see also Section 14).

In further work with James Bardeen and Carter, the analogy with the laws of thermodynamics was completed (\ref{SWH33}). The main point of the paper is to return to an examination of  what happens when matter or gravitational radiation falls into a black hole. They derived the first law of black holes mechanics, namely
\begin{equation}
{\rm d}M = \frac{\kappa\,{\rm d}A}{8\pi} + {\Phi}\,{\rm d}Q + {\Omega}\,{\rm d}J\ ,
\end{equation}
where $\kappa$ is the surface gravity of the black hole, a measure of the strength of the gravitational field at the event horizon, ${\rm d}A$ is the change in the area of the black hole, ${\Omega}$ is the angular velocity of the black hole, ${\rm d}J$ is the change in the angular momentum, $\Phi$ the electrostatic potential, ${\rm d}Q$ is the change in the charge of the black hole and ${\rm d}M$ the change in its mass. Were it the case that ${\kappa\,{\rm d}A/8\pi}$ were replaced by $T\,{\rm d}S$, $T$ being the temperature of the system and ${\rm d}S$ the change in its entropy, then relation (1) would be identical to the first law of thermodynamics that expresses the change in the energy of a system in terms of its heat content and the work done on the system. The analogy was bolstered by the observation that $\kappa$ did not depend on where on the horizon it was computed. This conclusion is much the same as the zeroth law of thermodynamics which states that if a body is in thermal equilibrium then its temperature is independent of position. The final observation was that the limit $\kappa \rightarrow 0$ could not be achieved in any finite process, loosely equivalent to the third law of thermodynamics that reaching absolute zero is an impossibility.

\section{Primordial Black Holes}

Stephen was one of the first to realise that black holes could form in the early universe due to the great compression of the Big Bang. Such primordial black holes (PBHs) would have the particle horizon mass at formation and could form with masses all the way down to the Planck mass, $2 \times 10^{-5}$ g. His first paper on the subject in 1971 was motivated by the possibility that tiny PBHs could be electrically charged and capture electrons to form `atoms' (\ref{Hawking1971}). They could then be detected in bubble chambers or captured by stars. With his later discovery of black hole evaporation, the model was invalidated but this work essentially started the field.

In fact, Zel'dovich and Igor Novikov had also considered PBHs in 1967 but they had come to a rather negative conclusion \citep{zeldovichnovikov1967}. A simple Bondi accretion analysis suggested that PBHs would grow as fast as the universe throughout the radiation-dominated era and thus attain a mass of order $10^{15}\,M_{\odot}$ by today. The existence of such huge black holes could be excluded, which suggested that PBHs never formed. However, their argument was Newtonian and neglected the cosmological expansion.  In 1974 Carr and Hawking showed that there is no self-similar solution in which a black hole grows as fast as the universe (\ref{carrhawking1974}). There was therefore no reason to discount PBHs and perhaps the most important consequence was that it motivated Stephen to consider quantum effects in small black holes.

The most natural mechanism for PBH formation involves the collapse of primordial inhomogeneities \citep{Carr:1975qj}, such as might arise from inflation.  Stephen also explored other possibilities, however, such as the collision of bubbles of broken symmetry with Stewart and Ian Moss (\ref{hawkingmossstewart1982}) and the collapse of cosmic strings (\ref{Hawking1989}). Stephen's most extensive work on PBH formation was a series of papers with Raphael Bousso on PBH production during inflation from gravitational instanton effects in Euclidean quantum gravity (\ref{BussoHawking1995}, \ref{BussoHawking1997}).

Interest in PBHs intensified when Stephen discovered that black holes radiate, since only PBHs could be small enough for this to be important (Section 9). In particular, those PBHs with masses less than $10^{15}$ g would have evaporated by the present epoch. The radiation from a solar-mass black hole would have a temperature of only $10^{-7}$ K and so would be very difficult to detect. However, $10^{15}$ g PBHs would have a temperature of $10^{12}$ K and would terminate their evaporation at the present epoch in tremendous explosions of high-energy gamma rays. How powerful these explosions would be depends on how many different species of elementary particles there are. In the quark model, the explosion would have an energy equivalent to $10^7$ megaton hydrogen bombs. On the other hand, an alternative theory of elementary particles, put forward by Rolf Hagedorn, suggested that there are an infinite number of elementary particles of ever higher mass, in which case the final explosion could be $10^5$ times more powerful.

In Stephen's paper with Don Page in 1976, the 100 MeV gamma-ray background generated by $10^{15}$ g PBHs was estimated and compared with measurements made by the SAS-2 satellite (\ref{PageHawking1976}). They showed that the average cosmological density of PBHs must be less than about 200 per cubic light-year, although the local density could be $10^6$ times higher if PBHs are concentrated in the halos of galaxies. This implied that the closest PBH would be at least as far away as Pluto. They concluded that it would be difficult to fly a detector large enough to detect PBH explosions. Later David Cline and colleagues suggested that some of the short-period gamma-ray bursts could be exploding black holes \citep{Cline:1996zg}. This model is not the main-stream view and it would require something strange to have happen at the QCD temperature of about $10^{12}$ K  above which quark matter was created, but at least it has testable consequences, such as an anisotropic spatial distribution and a correlation between the burst energy and duration.

After 50 years there is still no definite evidence for either evaporating or non-evaporating PBHs, but even their non-existence would give valuable information, since it would indicate that the early universe was very smooth. Indeed, numerous upper limits on the density of PBHs in various mass ranges already constrain models of the early universe. In recent years, however, it has been suggested that non-evaporating PBHs could provide the dark matter \citep{Carr:2016drx} or the black-hole coalescences detected by LIGO \citep{Abbott:2016blz} and even the supermassive holes which reside in galactic nuclei and which power quasars \citep{Carr:2018rid}. The topic has now become very popular and non-evaporating PBHs could turn out to play a more important cosmological role than evaporating ones. If so, Stephen's pioneering work on the topic may turn out to be one of his most important and prescient contributions.

\section{Black Hole Radiation}

By 1973, classical black hole theory had reached a degree of maturity. Black holes, however, contain space-time singularities, regions where general relativity breaks down, raising fundamental issues about the nature of space and time. Because of this breakdown at singularities, the next obvious step was to attempt to combine general relativity, the theory of the very large, with quantum theory, the theory of the very small.

In 1972, Jacob Bekenstein had suggested that a black hole has an entropy proportional to the area of its event horizon. Stephen was highly sceptical because black holes would then have a non-zero temperature. Any object with a temperature emits black body radiation, but black holes were thought to be unable to emit anything. Stephen began to study how particles and fields governed by quantum theory would be scattered by a black hole.   Earlier studies of super-radiance had show that an incoming wave could be amplified by scattering off a spinning black hole \citep{Zeldovich1971,Starobinsky:1973aij} and had even indicated, as an extension, that a spinning black hole could superradiantly scatter vacuum fluctuations, causing spontaneous particle emission that spins the black hole down.

At the Krakow meeting on Cosmology in September 1973, Stephen met Zel'dovich and Aleksander Starobinsky who pointed out to him this spontaneous emission \citep{Thorne1994}.   Stimulated by these discussions, Stephen discovered a few months later, much to everyone's surprise, that all black holes, even those that don't spin and thus have no spin energy to extract, emit particles at a steady rate. Initially, he thought there must be a mistake in his calculation. That black holes could emit particles went against the long-held belief that black holes could only be absorbers and never emitters of anything. The result refused to go away. What finally convinced Stephen was that the outgoing particles had precisely a thermal spectrum -- the black hole created and emitted particles and radiation just like any hot body with a temperature $T = \kappa/2\pi$ where $\kappa$ is the surface gravity of the black hole.  This phenomenon came to be known as \emph{Hawking radiation}.

For a non-rotating black hole, the temperature is given by Stephen's most famous equation,
\begin{equation}
T = \frac{\hbar c^3}{8\pi Gk_{\rm B}M}\ ,
\end{equation}
where $\hbar$ is the reduced Planck constant, $c$ the speed of light, $G$ Newton's gravitational constant and $k_{\rm B}$ is Boltzmann's constant. For astronomical black holes, the temperature is tiny, $T \sim 10^{-7}\,(M/{M_{\odot}})\,{\rm K}$, where $M_{\odot}$ is the mass of the Sun. It is fair to say that Stephen's discovery ranks as one of the most important results ever in fundamental physics. It means that the first law of black hole mechanics should be identified with the first law of thermodynamics,
\begin{equation}
{\rm d}M = T\,{\rm d}S + {\Phi}\,{\rm d}Q +{\Omega}\,{\rm d}J\ .
\end{equation}
Comparing the relations (1) and (3), the entropy $S$ of the black hole can be identified as,
\begin{equation}
S = \frac{Ak_{\rm B}c^3}{4G\hbar}\ ,
\end{equation}
a result confirming Bekenstein's intuition. Unlike the entropy of other systems, however, black hole entropy is not extensive, being proportional to the horizon's area $A$ and not its volume.

Consequently an isolated black hole emits thermal radiation and so loses mass. Since expression (2) shows that the temperature is inversely proportional to the mass, the temperature increases, leading to a runaway process and eventually, it is presumed, to the disappearance of the black hole. The process takes an inordinately long time for a solar mass black hole, about $10^{67}$ years. But for small black holes, which could have been formed in the early universe (see Section 8), their lifetimes could be much shorter. For example, a black hole of mass $10^{15}$ gm will last about $10^{10}$ years, approximately the current age of the universe.

Stephen's first picture for the radiation mechanism was that vacuum fluctuations in the region outside the black hole give rise to pairs of virtual particles. One member of the pair would have positive energy and escape to infinity, where it would appear as radiation. The other would have negative energy but it could continue to exist without having to annihilate because a black hole contains negative energy states. One can think of the negative-energy particle as a positive-energy particle travelling backwards in time from the black hole singularity until it is scattered forward in time where the virtual pair first appears.

Stephen first announced this result at a meeting on Quantum Gravity at the Rutherford Laboratory on 15--16 February 1974 and it was published in \emph{Nature} shortly afterwards (\ref{hawking1974}). The prediction has now been derived in several different ways and it is such a beautiful result -- unifying quantum theory, general relativity and thermodynamics -- that most experts accept that it is correct. Wheeler once said that just talking about it was like `rolling candy on the tongue'.

In his original paper, Stephen obtained the radiation by considering zero-point fluctuations in the initial vacuum state that were amplified by the collapse that formed the black hole. However, the fluctuations would receive a huge redshift just outside the horizon, so that the radiation seemed to come from modes that were initially far above the Planck frequency. The theory might break down at such high frequencies and so Stephen sought a mathematical treatment of black hole radiation as low energy particles leaking out of the horizon at late times, rather than as a high energy process during the collapse itself. Together with Hartle, he showed how this can be achieved by using path integrals to calculate the amplitude for a scalar particle to propagate in the curved spacetime of a black hole from the future singularity to an observer at infinity (\ref{HH76}). In order to make the path integral converge, it was necessary to complexify the spacetime, thereby connecting the past and future singularities of the black hole. The emission and absorption probabilities for a particle of energy $\varepsilon$ escaping from or falling into the black hole were related by the Boltzmann factor $\exp(-\varepsilon/T)$, which was precisely the relation needed for the black hole to be in equilibrium with thermal radiation at temperature $T$. This derivation of thermal radiation avoided the questionable use of frequencies above the Planck value and confirmed that the radiation corresponds to energy leaking out of the horizon at late times rather than during the collapse.

The paper with Hartle showed that the Schwarzschild solution can be analytically continued to a section on which it is Euclidean, that is, with a positive--definite metric. The natural choice of propagator was then the unique Green's function on this Euclidean section. When this propagator is analytically continued back to the Lorentzian Schwarzschild solution, it has poles periodic in the imaginary time coordinate. Gibbons and Malcolm Perry recognized this as the characteristic signature of thermal Green's functions, which meant that the proof of thermal emission could be extended to interacting field theories as well (\ref{gibbonshawkingperry1978}) (Figure 8).

This work showed that general relativity can be combined with quantum theory in an elegant manner by adopting the Euclidean approach in which ordinary time is replaced by imaginary time and becomes a fourth direction of space. Much of the early work on Euclidean quantum gravity was carried out with his Cambridge colleagues at that time -- Gibbons, Page and Christopher Pope (\ref{HawkingPagePope1980}) -- culminating in an edited volume on the topic (\ref{GibbonsHawking1991}). Later Stephen and Hartle extended this approach to cosmology with their `No Boundary' proposal (\ref{HH83}) (see Section 12).

Stephen eventually became pessimistic about seeing direct proof of Hawking radiation.  There are, however, solid state analogues of black holes and cyclotron effects that might be accepted as proof. Also there is another kind of Hawking radiation of much longer wavelength originating from the cosmological horizon of the early inflationary universe which might be detected as primordial gravitational waves. This possibility arises because in 1977, Gibbons and Stephen showed that there is a temperature associated with the horizon of the de Sitter model (\ref{SWH49}). If such primordial gravitational radiation were detected, then black holes almost certainly emit radiation. The work with Gibbons was also important because it led to an understanding of how density fluctuations arise in the inflationary model (\ref{HawkingMoss1983}) (see Section 12).

\section{First Concerns about the Information Paradox }

While visiting CalTech for the academic year 1974-75, Stephen became increasingly concerned about the significance of black holes for fundamental physics.  He wrote two papers that totally changed the way we think about black holes.   The first paper examined the thermodynamics of black holes by considering how a black hole can come into thermal equilibrium with a bath of radiation in the microcanonical ensemble (\ref{Hawking1976}).   But in thermal equilibrium, one cannot determine the direction of time by observation. In classical general relativity, a black hole is a region from which one cannot escape, but in the interior of which there is a space-time singularity.  According to the ideas of cosmic censorship, one need not worry about these singularities. They are unobservable to exterior observers as they are inside the event horizon of the black hole.

The time reverse of a black hole is a white hole. It has a singularity in the past and defines a region of space from which matter and radiation \emph{must} escape. The ideas of predictability and cosmic censorship mean that it is believed that such objects do not exist in nature. Stephen argued in the second paper that, unlike what happens according to classical physics, in quantum physics black holes and white holes must be indistinguishable to outside observers (\ref{SWH42}).  These ideas led him to two radical conclusions. The first was that a description of space-time will be dependent on what the observer is doing rather than having an objective existence.  The second conclusion, greatly elaborated in his paper \emph{Breakdown of Predictability in Gravitational Collapse}, is that gravitational physics requires the introduction of extra uncertainty into the fundamental laws of nature.

The basic argument that makes this new unpredictability clear is to think about how a black hole is formed. The black hole uniqueness theorems showed that the only properties that a classical black hole can have are its mass, electric charge and angular momentum. But a black hole can be formed from the gravitational collapse of any collection of matter. All that has to happen is that a sufficient amount of mass--energy occupies a sufficiently small region of space. If the spatial extent $R$ of a region of mass-energy $M$ has value $R < {2GM/c^2}$, then a black hole will be formed.  The nature of the black hole appears to be independent of its mode of formation. Eventually, the black hole evaporates producing nothing but thermal Hawking radiation. As it loses mass, it gets hotter and hotter and eventually evaporates completely. Stephen proposed that such a process was governed by a superscattering operator $\$$ which should be thought of as a generalisation of the quantum-mechanical S-matrix and would require a generalisation of quantum mechanics.  The S-matrix maps initial quantum states into final quantum states. Instead, the $\$$ operator maps initial density matrices into final density matrices.  Were black hole physics to work in this way, it would be in conflict with quantum mechanics.

To see this, imagine that the initial matter configuration was constructed from a pure quantum state, in technical terms, a state with vanishing von Neumann entropy. The final thermal radiation produced by the black hole has huge von Neumann entropy and is independent of how the black hole was formed.  Quantum mechanics requires the time evolution of a system to be unitary, and consequently the von Neumann entropy is constant in time. The Hawking radiation produced by the black hole seems to be independent of how the black hole  was formed. This conflict between quantum mechanics and the semi-classical picture of black hole evaporation is the \emph{information paradox}.

Although Stephen agonised about the paradox on and off over the next forty years, it was only in his very last works that a solution began to appear (Section 14).

\section{Topology Change, Acausality and Wormholes}

After the successes of the path integral approach to black hole thermodynamics, Stephen began exploring its wider implications.  Adopting the graphic style and some of the ideas of Wheeler, he envisaged the vacuum as having a foam-like structure, the bubbles being associated with Euclidean, or more precisely, Riemannian solutions of the Einstein
equations with non-trivial topology. By analogy with contemporary developments in Yang-Mills theory of elementary particles, such non-singular metrics were dubbed `gravitational instantons' (\ref{SWH46}). Most of the known examples admitted at least one isometry whose fixed point sets were either zero or two dimensional. The first case occurs in the Taub-NUT solution given in Stephen's paper of 1977 (\ref{SWH46}) and the second in the Riemannian Schwarzschild solution.  In the classification scheme developed in his paper of 1979 with Gibbons, Stephen christened these, slightly whimsically, NUTS and BOLTS respectively (\ref{SWH57}). The former are a gravitational analogue of Dirac magnetic monopoles.  Of particular interest were solutions whose curvature tensor was self-dual, known to mathematicians as hyper-K\"ahler metrics. Stephen's Taub-NUT solution is self-dual as are its multi-Taub-NUT brethren (\ref{SWH60}).

The subject of gravitational instantons excited the interest of differential geometers and sparked off considerable activity on their part in this field.  Gravitational instantons continue to play an important role in Kaluza-Klein theories, higher dimensional supergravity theories and superstring approaches to quantum gravity. Stephen himself was keenly interested in their possible effects on the propagation of elementary particles through a space-time foam seething with virtual black holes,  bubbles,  wormholes  and baby universes, as part of his belief at that time that unpredictability is an inevitable consequence of quantum gravity
(\ref{SWH42}, \ref{SWH47}, \ref{SWH50}, \ref{SWH58}, \ref{SWH59}, \ref{SWH60}, \ref{SWH62}, \ref{SWH63}, \ref{SWH92}, \ref{SWH108}, \ref{SWH118}, \ref{SWH155}, \ref{SWH166}).

Stephen questioned not only whether quantum gravity would be predictable, but also whether it would be causal (\ref{SWH66}, \ref{SWH67}, \ref{SWH150}). Suggested culprits were the `wormholes', possible entities whose potential importance had been emphasised by Wheeler. They can be thought of as short tunnels connecting apparently distant regions of space or space-time. In fact there are two different types of worm hole.  Those considered by Wheeler correspond to three-dimensional spatial sections of an ordinary four-dimensional space-time with non-trivial topology. An immediate question was whether dynamically the topology might change.  Wheeler's student Robert Geroch had shown, using what is called co-bordism theory that, although the possibility is allowed, any smooth metric in between two topologically distinct spatial sections must admit closed time-like curves which would therefore allow time travel. Much later Thorne argued that the mere existence of a wormhole could allow the construction of a time machine, although it had to be built of exotic material which violates the energy conditions used to establish the singularity theorems. Such violations are possible according to quantum field theory, but Stephen conjectured that the existence of closed time-like curves would lead to quantum field theoretic back reaction effects which prevent the construction of such wormholes in the first place (\ref{SWH134}, \ref{SWH169}).

Another objection to topology change, at least at the semi-classical level, involves spinor fields such as those which describe electrons, protons and other spin-half particles. While Geroch had shown that topology change is always possible in principle if one is prepared to admit closed time-like curves, there was a topological obstruction to the global existence of spinor fields (\ref{SWH136}, \ref{SWH138}). A space-time in which two such disjoint closed universes merged to form a single closed universe would be a very dangerous place for ordinary matter.

From the point of view of Euclidean quantum gravity, the relevant wormholes are 4-dimensional Riemannian manifolds with more than one asymptotic region. Stephen explored their possible effects in a series of papers (\ref{SWH121}, \ref{SWH127}, \ref{SWH129}, \ref{SWH132}).

\section{Nearly scale-invariant cosmological fluctuations -- the 1982 Nuffield Workshop}

Apart from a stream of influential journal publications and contributions to the proceedings of scientific conferences, Stephen, aided by Israel, commissioned articles from leaders in the field for two commemorative volumes, one to mark the centennial of Einstein's birth  (\ref{Hawking:1979ig}) and one to mark the tricentenary of the  publication of Newton's \emph{Principia Mathematica} (\ref{Hawking:1987en}). In addition he, together with colleagues in DAMTP, organised four workshops sponsored by the Nuffield Foundation in the years 1981, 1982, 1985 and 1989 on emerging areas of interest (\ref{Hawking:1981bu}, \ref{Gibbons:1984hx}, \ref{GHT}, \ref{Gibbons:1990gp}).

By far the most influential was the three-week workshop from June 21st to July 9th 1982 funded by the Nuffield Foundation which took place just less than a year after the publication of Alan Guth's influential paper on the \emph{inflationary scenario} \citep{Guth:1980zm}.  By that time the shortcomings of Guth's original scenario had emerged (\ref{Hawking:1981fz}, \ref{hawkingmossstewart1982}) \citep{Guth:1982pn} and a \emph{new inflationary scenario} proposed by Andrei Linde \citeyearpar{Linde:1981mu} and Andreas Albrecht and Paul Steinhardt \citeyearpar{Albrecht:1982wi} had replaced it.  The main outstanding issues addressed at the workshop concerned quantum fluctuations generated during the inflationary era of the universe and what preceded inflation.

Stephen had already made considerable headway in understanding this key issue for the formation of galaxies and the large-scale structure of the universe.   There is a close similarity between the cosmological and black hole event horizons.  In a de Sitter universe dominated by a cosmological constant, everything is simply turned inside out; observers surrounded by the cosmological event horizon find themselves immersed in thermal radiation (\ref{SWH49}).  In a natural progression from this work, Stephen played a key role developing the theory of inflationary fluctuations, for which there is now excellent observational evidence.

In spring 1982, Stephen visited the USA making a bold proposal about the quantum origin of galaxies, which was summarised in a preprint dated June of that year, but actually written before his trip to the USA (\ref{Hawking:1982cz}). During the de Sitter-like inflation era, he observed that the quantum fluctuations generated by this mechanism would have the right scale-invariant spectrum to explain the observations of large-scale structures in the universe today. Its origin was essentially Hawking radiation from the cosmological horizon, rather than the black hole horizon.  The amplitude of these fluctuations was subsequently worked out during the 1982 Nuffield workshop.

Among those who struggled with the problem of inflationary quantum perturbations were the authors of a number of contributions to the proceedings (\ref{HawkingMoss1983}, \ref{Hawking:1982cz}) \citep{Starobinsky:1982ee,Guth:1982ec,Bardeen:1983qw}. It is now widely acknowledged that these calculations and discussions at the Nuffield workshop formed the foundation of all subsequent work on inflationary perturbations.  In summary, fluctuations are expected to be produced by quantum fluctuations of any scalar fields and of the gravitational field in the de Sitter invariant vacuum, or ground state, to which the system would have settled down according to the `no-hair property' of the background predicted by Stephen, Gibbons and Moss (\ref{SWH49}, \ref{Hawking:1981fz}). Their properties had been elucidated in some detail by Bunch and Davies \citeyearpar{Bunch:1978yq}.

Hawking's seminars and preprint with the conceptual idea and the spectrum preceded these endeavours and the basic mechanism relies entirely on the breakthroughs that Stephen had made in understanding quantum fields around black holes as far back as 1974 and then together with Gibbons in de Sitter space in 1977. It subsequently emerged that a similar scale-invariant fluctuation result in the context of Starobinsky's model of the early universe was also obtained by Mukhanov and Chibisov working independently in the Soviet Union in 1981, although this was not known to Stephen and other Western researchers at that time \citep{MukhanovChibisov1981}.

In his conference summary at the end of the workshop, Frank Wilczek could state that
\begin{quote}
`Beautiful work on the spectrum of fluctuations expected in detailed inflationary models was carried on by several groups at the workshop, and mutual agreement was obtained after some struggles.' (\ref{Gibbons:1984hx}, p.\,477)
\end{quote}

The spectrum of the fluctuations was expected to be approximately scale-invariant in accordance with earlier ideas by Harrison \citeyearpar{Harrison:1969fb} and Zel'dovich \citeyearpar{Zeldovich}.   In the conclusion of his introduction to the proceedings (\ref{Gibbons:1984hx}), Stephen clearly revealed his Popperian credentials, stating that:
\begin{quote}
`The inflationary hypothesis has the great advantage that it makes predictions about the present density of the universe and about the spectrum of departures from spatial uniformity. It should be possible to test these in the fairly near future and either falsify the hypothesis, or strengthen it.'
\end{quote}

Despite all the odds, Stephen lived to see the observations of the Cosmic Microwave Background Radiation carried out in a spectacular fashion by the COBE (1989-1993), WMAP (2001-2011) and PLANCK (2009-2013) space observatories and to make that judgement to his own satisfaction.  None of the results of these missions is inconsistent with the broad picture of scalar fluctuations sketched out at the Nuffield workshop. The `smoking gun' for inflation is widely believed to lie with the tensor fluctuations, that is, with primordial gravitational waves.  Their detection is our current best bet for detecting Hawking radiation.

From 1982 onwards, Stephen concentrated his efforts on the deeper puzzle of the boundary conditions required to bring about inflation and the probability of them coming about. He tackled the issues at both the classical and the quantum level.  At the classical level the problem reduces to constructing a measure on a finite dimensional sub-space of the classical solutions of the Einstein equations (\ref{Gibbons:1986xk}). Difficulties arise because typically the total measure diverges and additional priors must be introduced (\ref{Hawking:1987bi}). The more ambitious quantum case is described in the next section.

\section{The Wavefunction of the Universe}

Stephen sought to understand the whole universe in scientific terms. As he said famously,
\begin{quotation}
`My goal is simple. It is a complete understanding of the universe.'
\end{quotation}

The singularity theorems proved by Stephen, Penrose, and others showed conclusively that the  classical Einstein equation implied that the universe began in a hot big bang. But the singularity theorems also showed that the beginning could not be described by  a classical  space-time geometry obeying the Einstein equation with three space and one time direction at each point. Rather they showed something more sweeping:  the classical Einstein equation breaks down at the big bang and along with that the notion that it could be described by a classical space-time.

The classical extrapolation into the past showed that, near the big bang, energy scales would have been reached at which the space-time geometry fluctuates quantum mechanically without a definite value -- quantum gravity.  As discussed in Section 9, earlier work of Stephen with Hartle demonstrated the power of Euclidean geometry to help understand the quantum Hawking radiation from evaporating black holes  (\ref{HH76}). It was therefore natural to try to use similar techniques to describe the quantum birth of the universe. Stephen first put forward a proposal along these lines at a conference in the Vatican in 1981, where he suggested that the universe began with a regular Euclidean geometry having four space dimensions that made a quantum transition to a Lorentzian geometry with three space and one time dimension that we have today (\ref{Vat81}).

To put this idea on a solid footing required a quantum state --- a wave function of the universe. Stephen and Hartle realized that for closed cosmologies this could be the cosmological analogue of the ground state constructed as a Euclidean functional integral (\ref{HH83}). The integral would be over geometries on a four-disk that matched the arguments of the wave function on its one boundary and were regular inside -- no other boundary.   Thus the no-boundary wave function, or the no boundary proposal, was born.

In the early 1980's Stephen, building on his work with Gibbons on quantum field theory in de Sitter spacetime (\ref{SWH49}), showed that inhomogeneities in the early universe could have arisen from quantum vacuum fluctuations that grew during an early period of inflation and eventually collapsed under gravity to form the large-scale structures we observe today (\ref{Hawking:1982cz}). Following the successful Nuffield Workshop described in Section 11 (\ref{Gibbons:1984hx}), subsequent observations confirmed the resulting predictions which must count as one of the great triumphs of theoretical cosmology, connecting the universe's earliest quantum evolution to the matter  distribution today. As a consequence of the growth of fluctuations, the universe also exhibits arrows of time such as that characterized by the global increase in thermodynamic entropy.

But how did inflation start and what selects a particular realization of inflation?   Stephen realized that early on there was a profound connection between the no-boundary wave function and inflation (\ref{HH83}, \ref{Hawking:1983hj}). In a series of papers over many years he and his collaborators consolidated this connection, showing that the no boundary proposal \emph{predicts} an early period of inflation. Specifically, Stephen and Jonathan Halliwell showed that the no boundary proposal describes an ensemble of universes in which inflation triggers the emergence of a classical Lorentzian space-time from the quantum fuzz at the beginning (\ref{Hartle:2008ng}), along with fluctuations that are initially predicted to be in their ground state (\ref{Halliwell:1984eu}).  Thus the no boundary proposal provides a foundation for inflationary cosmology.

Many successful cosmological predictions had been made using quantum fields assuming classical background space-times. But classical behavior is not a given in a quantum universe. Rather it is a matter of quantum probabilities. The no boundary proposal does not posit classical backgrounds -- it predicts them quantum mechanically, providing a unified origin for both classical backgrounds and quantum fluctuations -- a remarkable, simple, and beautiful achievement.

The scientific importance of the no boundary proposal is not just as a successful theory of the origin of the basic structure of the universe -- it also has had a significant impact on how we think about the universe and our place in it.

\begin{itemize}
\item\emph {A Quantum universe.} Quantum mechanics had been applied to cosmological models before, but the no boundary proposal made it inescapable that the universe is a quantum mechanical system whose observable properties follow from quantum mechanical probabilities, resulting in a renaissance of the field of quantum cosmology.

\item\emph{A new understanding of arrows of time.}  The no boundary proposal leads to an arrow of time that is initially aligned with the expansion, because it predicts that fluctuations start out in their ground state. Stephen first argued that the arrow of time would reverse in universes that contract after a period of expansion (\ref{Hawking:1985af}). However, after discussions with Page \citeyearpar{page1985}, he later worked with Raymond Laflamme and Glenn Lyons to show that the fluctuations in the no-boundary state continued to grow in a period of contraction, thereby giving rise to a thermodynamic arrow of time that points in a constant direction while the universe expands and contracts again (\ref{Hawking:1993tu}).

\item\emph{A new view of our role as observers --- top-down.} In a quantum theory of the universe observers are physical systems within the universe, not somehow outside it. The existence of observers selects a subset of the histories in the multiverse, meaning a  hypothetical group of multiple universes including our own universe, predicted by the no boundary proposal. Stephen was fond of this flexible notion of cosmological history that emerges from quantum cosmology.
\begin{quote}
`The history of the universe depends on the question we ask,'
\end{quote}
he used to say. He liked to call this a top-down approach to cosmology, reconstructing the universe's history starting from our position within it (\ref{Hawking:2003bf}, \ref{Hawking:2006ur}). The theoretical framework of quantum cosmology and the no boundary proposal imply a form of `anthropic' reasoning, but without the need to augment the theory with a separate anthropic principle.

The top-down approach has an important effect on the no boundary proposal predictions of inflation -- it is strongly biased towards a low level of inflation. Probabilities conditioned by our observational situation, however, predict a long period of inflation in our past (\ref{Hartle:2007gi}). As physical systems within the universe there is only a very, very small quantum probability for systems like us to have evolved in any region of a given size. It is therefore more probable that we live in a large universe generated by significant inflation because there are more places in which we could have evolved.

\item\emph{New formulations of Quantum Mechanics.} Stephen was not much interested in foundational issues in quantum mechanics. As he put it at least once
\begin{quote}
`When I hear the words `Schr\"odinger's Cat', I reach for my gun.'
\end{quote}
He thought we understood quantum mechanics well enough and, indeed, he was successful in applying it without worrying about any foundational questions.

But the no boundary proposal motivated new formulations of quantum mechanics that were adequate for cosmology -- decoherent histories, in particular. The usual textbook (Copenhagen) formulations of quantum mechanics are inadequate for cosmology not least because they predict probabilities of measurements made by observers. But in the very early universe no measurements were being made and there were no observers around to make them. A formulation of quantum mechanics general enough for cosmology was started by Hugh Everett and developed by many. That effort led to the decoherent (or consistent) histories approach to quantum theory and is adequate for quantum cosmology. It implies however that the Copenhagen interpretation of quantum theory is an approximation for measurement situations.

\item\emph{A new view of a `final theory'.} Including a wave function of the universe means that final theories consist of two parts ---  a theory of the universe's dynamics and a theory of the wave function of the universe, which are potentially unified as in the no boundary proposal. There are no predictions of any kind that do not involve both at some level.
\end{itemize}

In his last work in cosmology, Stephen, Hartle and Hertog showed that the histories dominating the top-down probabilities in the no boundary proposal have a regime of so-called eternal inflation where the quantum effects dominate the universe's evolution (\ref{Hartle:2010dq}) (Figure 9). This appears to spread out the wave function over a vast or even an infinite number of different kinds of inflationary universes, leading to the popular view that globally the universe would develop like a fractal consisting of infinitely many pocket universes separated from each other by an eternally inflating ocean.

Stephen viewed this as a breakdown of the semi-classical theory rather than a firm theoretical prediction, arguing that one had to go beyond the semi-classical approximation to the no boundary proposal to describe properly eternal inflation.  To do so, however, required new developments in quantum gravity, which eventually came in the form of holography. As discussed in Section 14, holography provided a realization of the idea that our visible universe may be a four-dimensional membrane in a higher-dimensional space. With Andrew Chamblin and Harvey Reall, Stephen studied black holes formed by matter trapped on such membranes (\ref{Chamblin:1999by}). In a Euclidean setting, holography provided a new application basis for Stephen's Euclidean quantum gravity programme, the highlights of which had meanwhile been brought together in a single volume (\ref{Gibbons:1994cg}). In a paper \emph{Brane New World}, Stephen, working with Hertog and Reall, described the creation of such membrane cosmologies in the context of the no boundary proposal (\ref{Hawking:2000kj}). They also drew on holographic techniques to refine the predictions of the spectral properties of inflationary fluctuations (\ref{Hawking:2000bb}).

Working with Hartle and Hertog, Stephen realized that holography enabled a new formulation of the no boundary proposal in which the dimension of time is holographic and projected out rather than transformed into a space dimension (\ref{Hartle:2012tv}). Stephen and his collaborators then put forward a holographic model of eternal inflation in the no-boundary theory in which the quantum regime of eternal inflation lives on a past boundary surface, a rather radical departure from the original no-boundary idea.  In his last paper on cosmology, Stephen argued that the holographic form of the wave function reduces the multiverse in eternal inflation to a manageable set of largely uniform and finite universes, giving him the grip on the multiverse he had always searched for (\ref{Hawking:2017wrd}).

Stephen always called the theory a `proposal' for the quantum beginning of the universe. We have yet to see whether its predictions agree with future observations and, if so, whether it is unique in some sense.  Stephen is also on record as regarding the no boundary proposal as his best achievement in science. His vision to bring the question of the boundary conditions of the universe firmly within the realm of the physical sciences and his relentless pursuit of a simple, manageable quantum theory of the beginning constitute a giant conceptual leap forward,  whether the no boundary proposal proves to be correct or not.

\section{The Information Paradox Revisited}

Over the forty years since its discovery, there have been many suggestions concerning how the information paradox discussed in Section 10 could be resolved.  Stephen vacillated over the issue for many years, from believing in this loss of information on the one hand to trying to rescue the quantum mechanical picture on the other.   Eventually he became convinced that the quantum mechanical picture is correct, finally being persuaded by the anti-de Sitter space -- Conformal Field Theory (AdS/CFT) correspondence discovered by Juan Maldacena \citeyearpar{1999IJTP...38.1113M}.  Anti-de Sitter space was originally conceived of as a cosmological model for a universe that has negative cosmological constant and no matter. Its spatial sections are hyperbolic three-space.  If one looks however at time-like geodesics in this spacetime, although space is infinite, the geodesics never reach spatial infinity but are re-focused back into the interior of the space. In this sense, anti-de Sitter space is like a box that confines all fields, including the gravitational field.

Maldacena's discovery was that gravitation in anti-de Sitter space is equivalent to a non-gravitational quantum field theory defined on the boundary of anti-de Sitter space, that is, on the walls of the box. A question asked about gravity therefore becomes a question about a perfectly ordinary quantum field theory. Maldacena's discovery does not qualify as a proof, but more as an illustration of what is expected to be a general picture.   Therefore, there could be black holes in anti-de Sitter space that were described by perfectly ordinary quantum field theories on the boundary. If true, black holes must be described by quantum mechanics and no generalisation is required, whilst leaving unresolved the problem of explaining how this actually works and what is wrong with the semi-classical picture.

Thus, it might be that quantum mechanics holds good and information about the collapse is somehow encoded in the Hawking radiation, but that would require a revision of the black hole uniqueness theorems.  Remarkably, that turned out to be the case. With Perry and Andrew Strominger, Stephen found that the no-hair theorems are in need of modification to take account of what is known as `soft hair' (Figure 10).

Wheeler, following Bekenstein, characterised the physics of black holes by the aphorism `black holes have no hair'.  What he meant is that it is very hard to tell black holes apart because there are only three macroscopic quantities that describe them: their mass, angular momentum and electric charge. Wheeler felt that, since he could tell people apart by looking at their hair, having no hair was a good analogy. Black holes to him were indistinguishable from one another -- you could not even tell if a black hole had originated in a star made of matter or antimatter.  But if you look closely at bald people, you will find that they have small soft hairs close to the scalp that are hard to see at a distance. The same is true of the event horizon of a black hole which also has soft hair. In previous studies, soft hairs were ignored because they were thought to be fake degrees of freedom of the gravitational or electromagnetic fields -- they were termed `soft charges'.   They turn out however to be pure gauge degrees of freedom that are \emph{not} physically redundant. They provide a way of making black holes distinguishable. One can also think of soft hair as being black hole analogues of gravitational memory.

Gravitational memory is usually thought of as being something associated with the Bondi-Metzner-Sachs (BMS) group. One usually thinks of the symmetries of Minkowski space as being the Poincar\'e group, expressing the fact that flat Minkowski space does not have a preferred origin, a preferred direction or any notion of absolute velocity. These symmetries, when combined with the idea that the velocity of light is the same for all observers, contain the entire content of the special theory of relativity.  If one is sufficiently far away from any object, its gravitational field is expected to be so weak that for the most part it can be ignored. One might have expected the Poincar\'e group still to represent the symmetries of spacetime, but that expectation is wrong, as was first shown by Hermann Bondi (FRS 1959), Kenneth Metzner and Sachs. There is a much larger group of symmetries at large distances from isolated gravitating objects, the BMS group.

BMS transformations arise from gravitational radiation passing through a system. One might think that space-time would be the same before and after gravitational radiation has passed through the system, but there are subtle differences.  For example, in a LIGO type interferometer, the two mirrors might undergo a permanent displacement so that their separations before and after the passage of the gravitational wave are different. This is known as the memory effect. It arises from a pure gauge transformation of the gravitational field, but nevertheless, one that is observable.

Soft hair is precisely analogous to this memory effect, except that, instead of being far away from a gravitating object, it can be found on the event horizon of the black hole itself. Black holes have soft hair that make them distinguishable from one another and give them an infinite collection of extra properties that are in principle observable (\ref{HawkingPerryStrominger2016}). These ideas were further developed in a later paper in which  soft hair was used to calculate the entropy of black holes (\ref{BacoHawkingPerryStrominger2018}). The entropy arises from a holographic conformal field theory on the horizon of the black holes, microstates of which reproduce the black hole entropy.  In the Boltzmann interpretation of entropy, these microstates determine the quantum state of the black hole. Thus, the black hole could have its quantum state determined by how the black hole formed.

Whilst this is not a solution to the information paradox, it does pave the way for future work that might provide its complete resolution. The soft hair allows information from the formation of the black hole to be preserved. It remains to be determined if this is sufficient to rescue quantum mechanics.

\section{In Memorium}

We cherish the astonishing pictures of Stephen in 2007 `floating' weightlessly in the NASA `Pathfinder Flight' aircraft and manifestly overjoyed at escaping, albeit briefly, from the clutches of the gravitational force he had studied for decades and which had so cruelly imprisoned his body (Figure 11).

Stephen died on 14th March 2018, the 139th anniversary of the birth of Albert Einstein.   His ashes were interred in Westminster Abbey following a memorial service on 15th June 2018, the memorial stone being placed between the graves of Isaac Newton and Charles Darwin (FRS 1839).  On that stone, his equation for the temperature of a black hole is engraved while on a second memorial stone at Caius College his equation for the entropy of a black hole is displayed, as he requested (Figure 12).

\section*{Acknowledgements}

The editors are most grateful to the authors of this Memoir for their writings and their collaboration in producing a single coherent account of the Stephen's life and work. Malcolm Perry and Anna {\.Z}ytkow were particularly helpful in editing the Memoir and clarifying parts of the story.   We are also very grateful to Jane Hawking for her sensitive scrutiny of the Memoir and to the trustees of Stephen's Estate for permission to reproduce the early images of Stephen and their family.  Anna {\.Z}ytkov kindly provide colour images from her vast collection of photographs recording Stephen's activities.  The cover portrait shows Stephen on his 60th birthday in 2002 and is: \copyright Dr. Anna {\.Z}ytkow.

\section*{AWARDS AND RECOGNITION}

\begin{itemize}\itemsep=0pt
\item[]Adams Prize (1966)
\item[]Eddington Medal (1975)
\item[]Pius XI Medal (1975)
\item[]Maxwell Medal and Prize (1976)
\item[]Heineman Prize (1976)
\item[]Hughes Medal (1976)
\item[]Albert Einstein Award (1978)
\item[]RAS Gold Medal (1985)
\item[]Dirac Medal (1987)
\item[]Wolf Prize (1988)
\item[]Prince of Asturias Award (1989)
\item[]Andrew Gemant Award (1998)
\item[]Naylor Prize and Lectureship (1999)
\item[]Lilienfeld Prize (1999)
\item[]Albert Medal (1999)
\item[]Copley Medal (2006)
\item[]Presidential Medal of Freedom (2009)
\item[]Breakthrough Prize in Fundamental Physics (2012)
\item[]BBVA Foundation Frontiers of Knowledge Award (2015)
\end{itemize}

\newpage

\section*{References}

\begin{Hawkingpapers}

\Hbibitem{Hawking1965}(1965)
 On the Hoyle--Narlikar theory of gravitation,
  \emph{Proceedings of the Royal Society of London A} {\bf 286} 313--319.
  (doi: \doi{10.1098/rspa.1965.0146})

\Hbibitem{HawkingEllis1965}(1965)
 (With G.~F.~R.~Ellis) Singularities in homogeneous world models,
  \emph{Physics Letters} {\bf 17} 246--247. (doi: \doi{10.1016/0031-9163(65)90510-X})

\Hbibitem{Hawking1965a}(1965)
  S.~W.~Hawking,
 Occurrence of singularities in open universes,
  \emph{Physical Review Letters} {\bf 15} 689--690. (doi: \doi{10.1103/PhysRevLett.15.689})

\Hbibitem{Hawking1966a}(1966)
 Perturbations of an expanding universe,
  \emph{The Astrophysical Journal} {\bf 145} 544--554.
  (doi: \doi{10.1086/148793})

\Hbibitem{Hawking1966b}(1966)
 Singularities in the universe,
  \emph{Physical Review Letters} {\bf 17} 444--445.
  (doi: \doi{10.1103/PhysRevLett.17.444})

\Hbibitem{Hawking1966c}(1966)
 The Occurrence of singularities in cosmology,
 \emph{Proceedings of the Royal Society of London A} {\bf 294} 511--521.
  (doi: \doi{10.1098/rspa.1966.0221})

\Hbibitem{Hawking1966d}(1966)
 Singularities and the geometry of space-time,
  \emph{Cambridge University Adams Prize Essay}. Republished: \emph{European Physical Journal H} {\bf 39} 413--503.
  (doi: \doi{10.1140/epjh/e2014-50013-6})

\Hbibitem{Hawking1966e}(1966)
 The Occurrence of singularities in cosmology. II,
  \emph{Proceedings of the Royal Society of London A} {\bf 295} 490--493.
  (doi: \doi{10.1098/rspa.1966.0255})

\Hbibitem{Hawking1966f}(1966)
 Properties of Expanding Universes,
  \emph{PhD Thesis, University of Cambridge}. Available at: https://www.repository.cam.ac.uk/handle/1810/251038.
  (doi: \doi{10.17863/CAM.11283})

\Hbibitem{HawkingTayler1966}(1966)
  (With J.~R.~Tayler)
 Helium production in anisotropic big bang universes,
  \emph{Nature}, {\bf 209} 1278--1279. (doi: \doi{10.1038/2091278a0})

\Hbibitem{Hawking1967}(1967)
 The occurrence of singularities in cosmology. III. Causality and singularities,
  \emph{Proceedings of the Royal Society of London A} {\bf 300} 187--201.
  (doi: \doi{10.1098/rspa.1967.0164})

\Hbibitem{HawkingEllis1968}(1968)
 (With G.~F.~R.~Ellis) The cosmic black body radiation and the existence of singularities in our universe,
  \emph{The Astrophysical Journal} {\bf 152} 25--36.
  (doi: \doi{10.1086/149520})

\Hbibitem{Hawking1968}(1968)
 Gravitational radiation in an expanding universe,
  \emph{Journal of Mathematical Physics} {\bf 9} 598--604.
  (doi: \doi{10.1063/1.1664615})

\Hbibitem{Hawking1969}(1969)
 On the Rotation of the Universe,
  \emph{Monthly Notices of the Royal Astronomical Society} {\bf 142} 129-141.
  (doi: \doi{10.1093/mnras/142.2.129})

\Hbibitem{HawkingPenrose1970}(1970)
 (With R.~Penrose) The Singularities of gravitational collapse and cosmology,
  \emph{Proceedings of the Royal Society of London A} {\bf 314} 529-548. (doi: \doi{10.1098/rspa.1970.0021})

\Hbibitem{Hawking1971}(1971)
 Gravitationally collapsed objects of very low mass,
  \emph{Monthly Notices of the Royal Astronomical Society} {\bf 152} 75--78. (doi: \doi{10.1093/mnras/152.1.75})

\Hbibitem{SWH22}(1971)
 (With G.~W.~Gibbons)  Evidence for black holes in binary star systems,
  \emph{Nature} {\bf 232} 465--466.
  (doi: \doi{10.1038/232465a0})

\Hbibitem{SWH23}(1971)
 Gravitational radiation from colliding black holes,
  \emph{Physical Review Letters} {\bf 26} 1344--1346.
  (doi: \doi{10.1103/PhysRevLett.26.1344})

\Hbibitem{SWH25}(1971)
 (With G.~W.~Gibbons) Theory of the detection of short bursts of gravitational radiation,
  \emph{Physical Review  D} {\bf 4} 2191--2197.
  (doi: \doi{10.1103/PhysRevD.4.2191})

\Hbibitem{SWH24}(1972)
S.~W.~Hawking,
\emph{Communications in Mathematical Physics} {\bf 25} 152--166.
(doi: \doi{10.1007/BF01877517})

\Hbibitem{SWH26}(1972)
Black holes in the Brans-Dicke theory of gravitation.
\emph{Communications in Mathematical Physics} {\bf 25}  167--171.
(doi: \doi{10.1007/BF01877518})

\Hbibitem{SWH28}(1972)
(With J.~B.~Hartle) Energy and angular momentum flow into a black hole.
\emph{Communications in Mathematical Physics} {\bf 27}  283-290.
(doi: \doi{10.1007/BF01645515})

\Hbibitem{SWH29}(1972)
(With J.~B.~Hartle) Solutions of the Einstein-Maxwell equations with many black holes.
\emph{Communications in Mathematical Physics} {\bf 26} 87--101.
(doi: \doi{10.1007/BF01645696})

\Hbibitem{SWH27}(1972)
 Gravitational radiation - the theoretical aspect,
  \emph{Contemporary Physics} {\bf 13} 273--282.
  (doi: \doi{10.1080/00107517208205681})

\Hbibitem{CollinsHawking1973a}(1973)
(With C.~B.~Collins) Why is the Universe isotropic?,
  \emph{The Astrophysical Journal} {\bf 180} 317--334. (doi: \doi{10.1086/151965})

\Hbibitem{CollinsHawking1973b}(1973)
 (With C.~B.~Collins) The rotation and distortion of the universe,
 \emph{Monthly Notices of the Royal Astronomical Society} {\bf 162} 307--320. (doi: \doi{10.1093/mnras/162.4.307})

\Hbibitem{HawkingEllis1973c}(1973)
 (With G.~F.~R.~Ellis) \emph{The large scale structure of space-time}, Cambridge: Cambridge University Press

\Hbibitem{SWH33}(1973)
(With J.~M.~Bardeen and B.~Carter) The Four laws of black hole mechanics.
\emph{Communications in Mathematical Physics} {\bf 31}  161--170.
(doi: \doi{10.1007/BF01645742})

\Hbibitem{carrhawking1974}(1974)
 (With B.~J.~ Carr) Black holes in the early Universe,
  \emph{Monthly Notices of the Royal Astronomical Society} {\bf 168} 399--415. (doi: \doi{10.1093/mnras/168.2.399})

\Hbibitem{hawking1974}(1974)
 Black hole explosions?,
  \emph{Nature} {\bf 248} 30--31. (doi: \doi{10.1038/248030a0})

\Hbibitem{PageHawking1976}(1976)
(With D.~Page) Gamma rays from primordial black holes,
 \emph{The Astrophysical Journal}, {\bf 206} 1--7. (doi: \doi{10.1086/154350})

\Hbibitem{HH76}(1976)
(With J.~B.~Hartle) Path Integral Derivation of Black Hole Radiance,
  \emph{Physical Review D} {\bf 13} 2188--2203. (doi: \doi{10.1103/PhysRevD.13.2188})

\Hbibitem{Hawking1976}(1976)
Black Holes and Thermodynamics,
\emph{Physical Review D} {\bf 13} 191--197. (doi: \doi{10.1103/PhysRevD.13.191})

\Hbibitem{SWH42}(1976)
 Breakdown of Predictability in Gravitational Collapse,
  \emph{Physical Review  D} {\bf 14} 2460--2473.
  (doi: \doi{10.1103/PhysRevD.14.2460})

\Hbibitem{SWH46}(1977)
  Gravitational Instantons,
 \emph{Physics Letters A} {\bf 60} 81--83.
  (doi: \doi{10.1016/0375-9601(77)90386-3})

\Hbibitem{SWH49}(1977)
(With G.~W.~Gibbons) Cosmological Event Horizons, Thermodynamics, and Particle Creation,''
  \emph{Physical Review  D} {\bf 15} 2738--2751.
  (doi: \doi{10.1103/PhysRevD.15.2738})

\Hbibitem{SWH47}(1978)
 Black Holes and Unpredictability,
  \emph{Physics Bulletin} {\bf 29} 23--24.

\Hbibitem{SWH50}(1978)
 (With C.~N.~Pope) Generalized Spin Structures in Quantum Gravity,
 \emph{Physics Letters B} {\bf 73} 42--44.
  (doi: \doi{10.1016/0370-2693(78)90167-3})

\Hbibitem{SWH59}(1978)
 (With C.~N.~Pope)  Symmetry Breaking by Instantons in Supergravity,
  \emph{Nuclear Physics B} {\bf 146} 381--392.
  (doi: \doi{10.1016/0550-3213(78)90073-1})

\Hbibitem{SWH60}(1978)
(With G.~W.~Gibbons) Gravitational Multi-Instantons,
 \emph{Physics Letters B} {\bf 78} 430--432. (doi: \doi{10.1016/0370-2693(78)90478-1})

\Hbibitem{gibbonshawkingperry1978}(1978)
 (With G.~W.~Gibbons and M.~J.~Perry)  Path Integrals and the Indefiniteness of the Gravitational Action,
 \emph{Nuclear Physics B} {\bf 138} 141--150 (doi: \doi{10.1016/0550-3213(78)90161-X})

\Hbibitem{SWH57}(1979)
 (With G.~W.~Gibbons) Classification of Gravitational Instanton Symmetries,
  \emph{Communications in Mathematical Physics} {\bf 66} 291--310.
  (doi: \doi{10.1007/BF01197189})

\Hbibitem{SWH58}(1979)
 (With D.~N.~Page and C.~N.~Pope) The Propagation Of Particles In Space-time Foam,'
 \emph{Physics Letters B} {\bf 86} 175--178.
  (doi: \doi{10.1016/0370-2693(79)90812-8})

\Hbibitem{SWH62}(1979)
 (With C.~N.~Pope) {Yang-Mills} Instantons and the S Matrix,
 \emph{Nuclear Physics B} {\bf 161} 93--111.
  (doi: \doi{10.1016/0550-3213(79)90128-7})

\Hbibitem{Hawking:1979ig}(1979)
 (With W.~Israel (eds))
 \emph{General Relativity : An Einstein Centenary Survey}, 919 p
 Cambridge: Cambridge University Press

\Hbibitem{SWH63}(1980)
  (With  D.~N.~Page and C.~N.~Pope) Quantum Gravitational Bubbles,
  \emph{Nuclear Physics B} {\bf 170} (1980) 283--306.
  (doi: \doi{10.1016/0550-3213(80)90151-0})

\Hbibitem{SWH66}(1980)
 `Acausal Propagation in Quantum Gravity',
  In \emph{Oxford:  Proceedings of Quantum Gravity Conference 2}, 393--415

\Hbibitem{HawkingPagePope1980}(1980)
 (With D.~N.~Page and C.~N.~Pope) Quantum Gravitational Bubbles,
 \emph{Nuclear Physics B} {\bf 170} 283--306. (doi: \doi{10.1016/0550-3213(80)90151-0})

\Hbibitem{Hawking:1981bu}(1981)
(With M.~Rocek)  Superspace And Supergravity. \emph{Proceedings, Nuffield Workshop, Cambridge, UK, June 16 -- July 12, 1980}. 527p,
Cambridge: Cambridge University Press

\Hbibitem{SWH67}(1981)
  Interacting Quantum Fields Around a Black Hole,
  \emph{Communications in Mathematical Physics} {\bf 80} 421--442.
  (doi: \doi{10.1007/BF01208279})

\Hbibitem{hawkingmossstewart1982}(1982)
 (With I.~G. Moss and J.~M.~Stewart)  Bubble Collisions in the Very Early Universe,
  \emph{Physical Review D}, {\bf 26} 2681--2693.
  (doi: \doi{10.1103/PhysRevD.26.2681})

\Hbibitem{Vat81}(1982)
 The Boundary Conditions of the Universe,
  \emph{Pontifical Academy of Sciences, Scripta Varia} {\bf 48} 563--572.

\Hbibitem{Hawking:1981fz}(1982)
 (With I.~G.~Moss)  Supercooled Phase Transitions in the Very Early Universe,
 \emph{Physics Letters B} {\bf 110} 35. (doi: \doi{10.1016/0370-2693(82)90946-7})

\Hbibitem{Hawking:1982cz}(1982)
  The Development of Irregularities in a Single Bubble Inflationary Universe,
 \emph{Physics Letters B} {\bf 115} 295--297.  The DAMTP preprint of this paper is dated June 1982. (doi: \doi{10.1016/0370-2693(82)90373-2})

\Hbibitem{HawkingMoss1983}(1983)
 (With I.~G.~Moss) Fluctuations in the Inflationary Universe  \emph{Nuclear Physics B} {\bf 224} (1983) 180--192. (doi: \doi{10.1016/0550-3213(83)90319-X})

\Hbibitem{HH83}(1983)
 (With J.~B.~Hartle) Wave Function of the Universe,
 \emph{Physical Review  D} {\bf 28} 2960--2975. (doi: \doi{10.1103/PhysRevD.28.2960})

\Hbibitem{Gibbons:1984hx}(1983)
(With G.~W.~Gibbons and S.~T.~C.~Siklos (eds)) \emph{The Very Early Universe. Proceedings, Nuffield Workshop, Cambridge, UK, June 21 -- July 9, 1982},
 Cambridge: Cambridge University Press

\Hbibitem{Hawking:1983hj}(1984)
  The Quantum State of the Universe,
  \emph{Nuclear Physics B} {\bf 239} 257--276. (doi: \doi{10.1016/0550-3213(84)90093-2})

\Hbibitem{SWH92}(1984)
 Nontrivial Topologies in Quantum Gravity
  \emph{Nuclear Physics B} {\bf 244} 135--146. (doi: \doi{10.1016/0550-3213(84)90185-8})

\Hbibitem{Halliwell:1984eu}(1985)
 (With J.~J.~Halliwell) Origin of Structure in the Universe,
  \emph{Physical Review  D} {\bf 31} 1777--1791. (doi: \doi{10.1103/PhysRevD.31.1777})

\Hbibitem{Hawking:1985af}(1985)
  The Arrow of Time in Cosmology,
  \emph{Physical Review D} {\bf 32} 2489--2495. (doi: \doi{10.1103/PhysRevD.32.2489})

\Hbibitem{GHT}(1986)
 (With G.~W.~Gibbons and P.~K.~Townsend (eds))
 \emph{Supersymmetry and its applications: superstrings, anomalies and supergravity} 481p. Cambridge, UK, June 23 -- July 14, 1985. Cambridge: Cambridge University Press

\Hbibitem{Gibbons:1986xk}(1987)
(With G.~W.~Gibbons and J.~M.~Stewart) A Natural Measure on the Set of All Universes,
\emph{Nuclear Physics B} {\bf 281} 736. (doi: \doi{10.1016/0550-3213(87)90425-1})

\Hbibitem{SWH108}(1987)
 Quantum Coherence Down the Wormhole,
 \emph{Physics Letters B} {\bf 195} 337--343. (doi: \doi{10.1016/0370-2693(87)90028-1})

\Hbibitem{Hawking:1987en}(1987)
 (With W.~Israel (eds)) \emph{Three Hundred Years Of Gravitation} 684 p.
 Cambridge: Cambridge University Press

\Hbibitem{Hawking:1987bi}(1988)
(With D.~N.~Page) How probable is inflation?,
  \emph{Nuclear Physics B} {\bf 298} 789. (doi: \doi{10.1016/0550-3213(88)90008-9})

\Hbibitem{SWH118}(1988)
 Wormholes in Space-Time,
 \emph{Physical Review  D} {\bf 37} 904--910. (doi: \doi{10.1103/PhysRevD.37.904})

\Hbibitem{Hawking1989}(1989)
 Black Holes From Cosmic Strings,
  \emph{Physics Letters B} {\bf 231} 237--239. (doi: \doi{10.1016/0370-2693(89)90206-2})

\Hbibitem{Gibbons:1990gp}(1990)
(With G.~W.~Gibbons, S.~W.~Hawking and T.~Vachaspati (eds)) \emph{The Formation and evolution of cosmic strings. Proceedings, Workshop, Cambridge, UK, July 3 -- 7, 1989}. 542p, Cambridge: Cambridge University Press

\Hbibitem{SWH121}(1990)
 Do Wormholes Fix the Constants of Nature?,
  \emph{Nuclear Physics B} {\bf 335} 155--165. (doi: \doi{10.1016/0550-3213(90)90175-D})

\Hbibitem{SWH127}(1990)
 (With D.~N.~Page) The spectrum of wormholes,
  \emph{Physical Review  D} {\bf 42} 2655--2663. (doi: \doi{10.1103/PhysRevD.42.2655})

\Hbibitem{SWH129}(1991)
 The Effective action for wormholes,
  \emph{Nuclear Physics B} {\bf 363} 117--131. (doi: \doi{10.1016/0550-3213(91)90237-R})

\Hbibitem{SWH132}(1991)
  The Alpha parameters of wormholes,
  \emph{Physica Scripta T} {\bf 36} 222--227.
  (doi: \doi{10.1088/0031-8949/1991/T36/023})

\Hbibitem{GibbonsHawking1991}(1991)
 (With G.~W.~Gibbons (eds)) \emph{Euclidean quantum gravity}. Singapore: World Scientific Publishing.

\Hbibitem{SWH134}(1992)
 Chronology protection conjecture,
  \emph{Physical Review  D} {\bf 46} 603--611.
  (doi: \doi{10.1103/PhysRevD.46.603})

\Hbibitem{SWH136}(1992)
 (With G.~W.~Gibbons)  Selection rules for topology change,
  \emph{Communications in Mathematical Physics} {\bf 148} 345--352.
  (doi: \doi{10.1007/BF02100864})

\Hbibitem{SWH138}(1992)
  (With G.~W.~Gibbons) Kinks and topology change,
  \emph{Physical Review Letters} {\bf 69} 1719--1721.
  (doi: \doi{10.1103/PhysRevLett.69.1719})

\Hbibitem{Hawking:1993tu}(1993)
  (With R.~Laflamme and G.~W.~Lyons)  The Origin of time asymmetry,
 \emph{Physical Review  D} {\bf 47} 5342--5356. (doi: \doi{10.1103/PhysRevD.47.5342})

\Hbibitem{Gibbons:1994cg}(1993)
  (With G.~W.~Gibbons (eds)) \emph{Euclidean quantum gravity},
  Singapore, Singapore: World Scientific Publishing.

\Hbibitem{BussoHawking1995}(1995)
 (With R.~Busso) Probability for primordial black holes,
  \emph{Physical Review D}, {\bf 52} 5659--5664. (doi: \doi{10.1103/PhysRevD.52.5659})

\Hbibitem{SWH150}(1995)
Quantum coherence and closed time-like curves,
  \emph{Physical Review  D} {\bf 52} 5681--5686.
  (doi: \doi{10.1103/PhysRevD.52.5681})

\Hbibitem{SWH155}(1996)
 Virtual black holes,
  \emph{Physical Review  D} {\bf 53} 3099--3107.
  (doi: \doi{10.1103/PhysRevD.53.3099})

\Hbibitem{SWH166}(1997)
 (With S.~F.~Ross)  Loss of quantum coherence through scattering off virtual black holes,
  \emph{Physical Review  D} {\bf 56} 6403--6415.
  (doi: \doi{10.1103/PhysRevD.56.6403})

\Hbibitem{BussoHawking1997}(1997)
 (With R.~Busso) Black holes in inflation,
  \emph{Nuclear Physics Proceedings Supplement} {\bf 57} 201--205. (doi: \doi{10.1016/S0920-5632(97)00377-0})

\Hbibitem{SWH169}(1998)
(With M.~J.~Cassidy) Models for chronology selection,
  \emph{Physical Review  D} {\bf 57} 2372--2380.
  (doi: \doi{10.1103/PhysRevD.57.2372})

\Hbibitem{Hawking:2000kj}(2000)
  (With T.~Hertog and H.~S.~Reall)   Brane new world,
  \emph{Physical Review  D} {\bf 62} 043501. (doi: \doi{10.1103/PhysRevD.62.043501})

\Hbibitem{Chamblin:1999by}(2000)
 (With A.~Chamblin and H.~S.~Reall)  Brane world black holes,
  \emph{Physical Review  D} {\bf 61} 065007. (doi: \doi{10.1103/PhysRevD.61.065007})

\Hbibitem{SWH189}(2000)
 (With T.~Hertog and N.~Turok) Gravitational waves in open de Sitter space,
  \emph{Physical Review  D} {\bf 62} 063502
  (doi: \doi{10.1103/PhysRevD.62.063502})

\Hbibitem{Hawking:2000bb}(2001)
  (With T.~Hertog and H.~S.~Reall)  Trace anomaly driven inflation,
  \emph{Physical Review  D} {\bf 63} 083504. (doi: \doi{10.1103/PhysRevD.63.083504})

\Hbibitem{Hawking:2003bf}(2003)
  Cosmology from the top down,
  In Bernard Carr: \emph{Universe or multiverse?} 91--98.

\Hbibitem{Hawking:2006ur}(2006)
  (With T.~Hertog) Populating the landscape: A Top down approach,
  \emph{Physical Review  D} {\bf 73} 123527. (doi: \doi{10.1103/PhysRevD.73.123527})

\Hbibitem{Hartle:2007gi}(2008)
(With J.~B.~Hartle and T.~Hertog)  No-Boundary Measure of the Universe,
  \emph{Physical Review Letters} {\bf 100} 201301. (doi: \doi{10.1103/PhysRevLett.100.201301})

\Hbibitem{Hartle:2008ng}(2008)
 (With J.~B.~Hartle and T.~Hertog)  The Classical Universes of the No-Boundary Quantum State,
  \emph{Physical Review  D} {\bf 77} 123537. (doi: \doi{10.1103/PhysRevD.77.123537})

\Hbibitem{Hartle:2010dq}(2011)
  (With J.~B.~Hartle and T.~Hertog)  Local Observation in Eternal inflation,
  \emph{Physical Review Letters} {\bf 106} 141302. (doi: \doi{10.1103/PhysRevLett.106.141302})
  
\Hbibitem{Hawking2013}(2013)
 My Brief History: A Memoir. London: Bantam Press.

\Hbibitem{Hartle:2012tv}(2014)
  (With J.~B.~Hartle and T.~Hertog) Quantum Probabilities for Inflation from Holography,
  \emph{Journal of Cosmology and Astroparticle Physics} {\bf 2014} 15 pages.

\Hbibitem{HawkingPerryStrominger2016}(2016)
  (With M.~J.~Perry and A.~Strominger) Soft Hair on Black Holes,
  Physical Review Letters {\bf 116} 231301. (doi: \doi{10.1103/PhysRevLett.116.231301})

\Hbibitem{HawkingPerryStrominger2017}(2017)
  (With M.~J.~Perry and A.~Strominger)  Superrotation Charge and Supertranslation Hair on Black Holes,
Journal of High Energy Physics {\bf 1705} 161. (doi: \doi{10.1007/JHEP05(2017)161})

\Hbibitem{Hawking:2017wrd}(2018)
  (With T.~Hertog) A Smooth Exit from Eternal Inflation?,
  \emph{Journal of High Energy Physics} {\bf 4} article 147. (doi: \doi{10.1007/JHEP04(2018)147})

\Hbibitem{BacoHawkingPerryStrominger2018}(2018)
  (With S.~Haco, M.~J.~Perry and A.~Strominger) Black Hole Entropy and Soft Hair (arXiv:1810.01847 [hep-th])

\end{Hawkingpapers}

\renewcommand{\refname}{References to Other Authors}
\bibliographystyle{QCP_Master_doi}
\bibliography{Non-Hawking_Papers}

%

\end{document}